\documentclass[journal,comsoc]{IEEEtran}

\usepackage{pifont,amssymb,url}
\usepackage{multicol,mathrsfs}
\usepackage[usenames,dvipsnames,svgnames,table]{xcolor}
\usepackage{bm,bbm,mathrsfs,amssymb,pifont,amssymb,pifont,amsfonts,amsmath,stmaryrd,color,amssymb,graphics,graphicx,epstopdf}
\usepackage{amscd,graphicx,array,dsfont,texdraw,tikz,enumerate,multicol,multirow,verbatim}
\usetikzlibrary{arrows,shapes,shadows,positioning,automata,patterns}
\usetikzlibrary{matrix,arrows,calc,trees,decorations.pathmorphing,decorations.markings, decorations.pathreplacing}
\usepackage[utf8x]{inputenc}
\usepackage[T1]{fontenc}
\usepackage{booktabs}
\usepackage{algorithm,algorithmicx,algpseudocode} 
\usepackage{bm,amsmath,amssymb,amsthm,graphicx,subfigure,epstopdf}
\usepackage[comma,numbers,square,sort&compress]{natbib}
\usepackage{epstopdf,epic,subfigure,epsfig,arydshln,enumerate}
\usepackage{fancyhdr,lastpage,color,dsfont}
\usepackage{tikz}
\usetikzlibrary{positioning}

\newtheorem{lemma}{\bf \em{Lemma}}
\newtheorem{thm}{\bf \em{Theorem}}
\newtheorem{rem}{\bf \em{Remark}}
\newtheorem{problem}{\bf \em{Problem}}

\newtheorem{assu}{\bf \em{Assumption}}

\usepackage{textcomp}
\usepackage{arydshln} 
\def\BibTeX{{\rm B\kern-.05em{\sc i\kern-.025em b}\kern-.08em
    T\kern-.1667em\lower.7ex\hbox{E}\kern-.125emX}}
\allowdisplaybreaks

\begin{document}
\title{Distributed Observer and Controller Design for Linear Systems: A Separation-Based Approach
}
\author{Ganghui Cao and Xunyuan Yin
\thanks{This research is supported by the National Research Foundation, Singapore, and PUB, Singapore’s National Water Agency under its RIE2025 Urban Solutions and Sustainability (USS) (Water) Centre of Excellence (CoE) Programme, awarded to Nanyang Environment \& Water Research Institute (NEWRI), Nanyang Technological University (NTU), Singapore. This research is also supported by the Ministry of Education, Singapore, under its Academic Research Fund Tier 1 (RG95/24 and RG63/22). Any opinions, findings and conclusions or recommendations expressed in this material are those of the author(s) and do not reflect the views of the National Research Foundation, Singapore and PUB, Singapore’s National Water Agency. (Corresponding author: Xunyuan Yin.)}
\thanks{Ganghui Cao and Xunyuan Yin are with the Nanyang Environment \& Water Research Institute, Nanyang Technological University, 1 CleanTech Loop, 637141, Singapore (e-mail: ganghui.cao@ntu.edu.sg, xunyuan.yin@ntu.edu.sg).}
\thanks{Xunyuan Yin is also with the School of Chemistry, Chemical Engineering and Biotechnology, Nanyang Technological University, 62 Nanyang Drive, 637459, Singapore.}
}
\maketitle

\begin{abstract}
This paper investigates the problem of consensus-based distributed control of linear time-invariant multi-channel systems subject to unknown inputs.
A distributed observer-based control framework is proposed, within which observer nodes and controller nodes collaboratively perform state estimation and control tasks.
Consensus refers to a distributed cooperative mechanism by which each observer node compares its state estimate with those of neighboring nodes, and use the resulting discrepancies to update its own state estimate.
One key contribution of this work is to show that the distributed observers and the distributed controllers can be designed independently, which parallels the classical separation principle.
This separability within the distributed framework is enabled by a discontinuous consensus strategy and two adaptive algorithms developed specifically for handling the unknown inputs.
Theoretical analysis and numerical simulation results demonstrate the effectiveness of the proposed framework in achieving state estimation, stabilization, and tracking control objectives. 
\end{abstract}

\begin{IEEEkeywords}
 Distributed state estimation, distributed control, multi-agent systems, consensus, sliding mode, adaptive systems.
\end{IEEEkeywords}

\section{Introduction}
\subsection{Separation Principle in Stabilization}\label{CSPIS}
Separation principle \cite{K.J.Astrom2021} provides a powerful way to design controllers for dynamical systems, whose full state is not directly measurable. Consider the following linear time-invariant system:
\begin{subequations}\label{classicLTI}
\begin{align}
	\dot x =& Ax + Bu \\
	y =& Cx,
\end{align}
\end{subequations}
where $x$, $u$, and $y$ are the state, input, and output, respectively. The most common observer for system \eqref{classicLTI} takes the following form \cite{K.J.Astrom2021}:
\begin{align} \label{classicObserver}
	\dot{\hat x} = A \hat{x} + Bu + L(C \hat{x} -y). 
\end{align}
Based on state estimate $\hat{x}$ in \eqref{classicObserver}, a linear feedback controller can be designed as
\begin{equation} \label{classicController}
	u=K\hat{x}.
\end{equation}
Combining \eqref{classicLTI} with \eqref{classicObserver} and \eqref{classicController}, the closed-loop system can be written as
\begin{align} \label{classicClosed}
	\begin{bmatrix} 
		\dot{x} \\
		\dot{\hat{x}}
	\end{bmatrix}
	= \begin{bmatrix} 
		A & BK\\
		-LC & A+BK+LC
	\end{bmatrix}
	\begin{bmatrix} 
		x \\
		\hat{x}
	\end{bmatrix}.
\end{align}
Define $e = x-\hat{x}$ as the state estimation error, and take nonsingular transformation 
\begin{equation*}
	\begin{bmatrix} 
		x \\
		e
	\end{bmatrix} =
	\begin{bmatrix} 
		I & 0 \\
		I & -I
	\end{bmatrix}
	\begin{bmatrix} 
		x \\
		\hat{x}
	\end{bmatrix}.
\end{equation*}
Then, \eqref{classicClosed} can be transformed into
\begin{align} \label{classicTansfored}
	\begin{bmatrix} 
		\dot{x} \\
		\dot{e}
	\end{bmatrix}
	= \begin{bmatrix} 
		A+BK & -BK\\
		0 & A+LC
	\end{bmatrix}
	\begin{bmatrix} 
		x \\
		e
	\end{bmatrix}.
\end{align}
From \eqref{classicTansfored}, it is concluded that the state of system \eqref{classicLTI} converges to zero, if and only if both $A+BK$ and $A+LC$ have eigenvalues
with negative real parts. This implies that system \eqref{classicLTI} can be stabilized by designing observer \eqref{classicObserver} and controller \eqref{classicController} separately.

\subsection{Distributed Stabilization Problem}\label{DSPformulation}
Modern industrial systems usually have a number of components or subsystems, and are typically equipped with multiple sensors and actuators. For such systems, there has been a spontaneous research interest in studying the distributed (also referred to as decentralized) stabilization problem \cite{Z.Gong1997,T.Kim2024}. The problem is formulated as follows. Consider a multi-channel linear system governed by
\begin{align*}
	\dot x =& Ax + \sum\limits_{i = 1}^{\bar N} {{B_i}{u_i}} \\
	{y_i} =& {C_i}x,\ i = 1,2, \cdots \bar N,
\end{align*}
where $\bar N$ agents are involved. The $i$th agent measures local output $y_i$ and applies local control input $u_i$ to the system. The control objective is to drive the system state to the origin. 

With the separation principle in mind, the first thought may well be designing controllers with the help of distributed observers. Recently developed consensus-based distributed observers \cite{F.C.Rego2019,Y.Pei2021} enable each agent to reconstruct the full state of the system. However, most distributed observers in the existing literature are designed for systems without inputs. When systems are subject to control inputs, those distributed observers fail to work unless each observer node has full access to the control inputs of the entire system. For case $\bar N = 3$, Fig.~\ref{inputdeli} illustrates how global control inputs are delivered to each observer node through a centralized communication mechanism. The limitations of this type of design are as follows:
\begin{figure}[htpb]\centering
	\centerline{\includegraphics[width=0.5\textwidth]{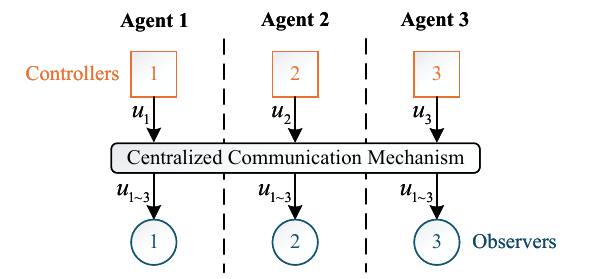}}
	\caption{Diagram of delivering all the control inputs to every observer node.}
	\label{inputdeli}
\end{figure} 
\begin{itemize}
	\item It is neither economic nor scalable. As $\bar N$ increases, the delivery of control inputs in real time from all controller nodes to every observer node becomes laborious and costly, leading to a substantial communication burden.
	\item It is fragile and unreliable. Since real-time delivery of global control inputs relies on a single centralized communication mechanism, any cyber attack or physical disruption on this communication mechanism may prevent observer nodes from receiving the complete control input information.
	\item It violates the distributed setup. Specifically, within a distributed framework, each agent is typically only allowed to communicate with a limited set of neighboring agents rather than with all others in the network (please see the recent works introduced in Section~\ref{CSTDS}). 
\end{itemize}
Due to the reasons discussed above, it is generally unwise to solve the distributed stabilization problem using distributed observers that rely on global input information. Instead, some dynamic output feedback frameworks are developed as alternatives, which are elaborated in the following.

\subsection{Current Solutions to Distributed Stabilization} \label{CSTDS}
\subsubsection{Conventional decentralized control}
In early research \cite{S.H.Wang1973}, each agent generates local control input $u_i$ only based on local output $y_i$. Due to the resulting information-structure constraints, the system cannot be stabilized if it has unstable fixed modes. To address this limitation, sampling controllers \cite{U.Ozguner1985,A.G.aghdam2008}, vibrational controllers \cite{T.Trave1985,S.Lee1987}, and periodically time-varying controllers \cite{B.Anderson1981,P.P.Khargonekar1994} have been explored over the decades. The authors in \cite{Z.Gong1997} gave a complete answer as to which fixed modes are removable and which are not, by using general nonlinear and time-varying decentralized controllers. By allowing agents to communicate with each other, the decentralized overlapping control framework was developed \cite{D.D.Vsiljak2005,J.Lavaei2008,J.Lavaei2009}.
This framework aims to characterize the necessary information that should be exchanged to achieve stabilization.
Optimal information exchange and minimum communication problems were investigated in \cite{S.Sojoudi2009,Y.Sturz2017}.
In addition, a recent interesting result shows that two classes of uncertain multi-variable linear systems are
stabilizable by a monotonically increasing diagonal gain matrix \cite{Z.Sun2021}.

\subsubsection{Consensus-based cooperative control}
Inspired by the success in distributed consensus control of multi-agent systems \cite{W.Ren2008book,Z.Li2017book}, recent studies have shown that the distributed stabilization problem can be more appropriately addressed using well-designed consensus protocols.
Within the consensus-based framework, agents are able to take full advantage of the information exchanged among them and act collaboratively. 
This enables the closed-loop system to bypass the aforementioned fixed-mode limitation and achieve improved control performance.

Pioneering work in \cite{K.Liu2017} and \cite{K.Liu2018} proposed consensus-based distributed observers to address cooperative stabilization and output-regulation problems, respectively. More precisely, dynamic output feedback control was investigated, while explicit state estimation was not pursued therein.
In \cite{X.Zhang2019}, the authors were concerned with systems that admit an appropriate block-diagonal decomposition, and proposed a distributed observer for either state estimation in the absence of control inputs, or dynamic output feedback stabilization.
Distributed state estimation and control problems were addressed in \cite{F.C.Rego2021} through a quantized, rate-limited consensus protocol. Based on this protocol, the difference between the state estimates from agents can be made sufficiently small, as the number of consensus iterations increases. This enables each agent to estimate the linear state feedback control inputs applied by other agents, which contributes to the convergence of both control and state estimation errors.
In another distributed stabilization framework, a consensus-based output estimator was developed, where each agent is responsible for estimating the global output of the entire system \cite{P.Duan2024TAC}. The output estimate is then fed into a state estimator residing in each agent. 
When agents apply linear state feedback control, co-designing the two estimators allows each agent to simultaneously achieve global input, global output, and full-state estimation.
A plug-and-play distributed stabilization framework was proposed in \cite{T.Kim2024}, where the stabilizing gains of the agents are computed by solving a Lyapunov equation in a fully distributed manner. In this framework, agents are allowed to join or leave the control loop freely during stabilization.
Recent advances presented in \cite{L.Wang2020ACC,F.Liu2023} also provide an insightful way to design distributed controllers based on the distributed observers developed in \cite{L.Wang2018}. It is shown that the distributed control problem can be reformulated as a conventional decentralized control problem, and subsequently be solved by resorting to the fundamental ideas and analysis techniques developed for the latter \cite{F.Liu2023}.
In addition, the information fusion strategies developed in \cite{S.P.Talebi2019,P.Duan2023TCNS,P.Duan2024TCNS,Y.Ren2024} show their unique capabilities in dealing with distributed optimal and privacy-preserving control problems.

\subsection{Common Limitations and Current Challenges}\label{CLACC}
Current consensus-based control frameworks share some common limitations and face challenges in some aspects:
\begin{itemize}
	\item There is no separation principle established in the above literature. For example, it has been explicitly pointed out in \cite{X.Zhang2019,F.C.Rego2021,P.Duan2024TCNS} that separation principle does not hold in the distributed setup. Consequently, co-design of the distributed controllers and observers becomes necessary, which substantially increases the complexity of the distributed stabilization problem.
	\item The distributed observers presented in previous studies (e.g., \cite{K.Liu2017,L.Wang2018,T.Kim2020}) typically ensure that the estimation errors converge to zero, when there are no control inputs or both the system states and control inputs approach zero. However, appropriately handling state estimation in the presence of
	persistent control inputs remains an open and ongoing research question.	
	\item The designed distributed controllers are not capable of stabilizing systems in the presence of unknown inputs that are often used to characterize system uncertainties~\cite{J.Chen1999book}. Moreover, only limited studies have considered distributed tracking control problems.
	\item The observer and controller design in an agent relies on the global output matrix ${\rm{col}}({C_{i}})_{i = 1}^{\bar N}$. If the $i$th agent only has access to local output matrix $C_i$, then additional consensus algorithms (such as Equation (16) in \cite{P.Duan2024TAC} and Equation (31) in \cite{T.Kim2024}) should be carried out to propagate the global output matrix or other concerned matrices over the agent network.
	\item As formulated in Section~\ref{DSPformulation}, an agent not only receives local output information, but also takes local control actions. In practice, however, a sensor and an actuator may not be located near each other and managed by the same agent. This motivates the need for a flexible framework that allows the distributed observers and controllers to be designed separately.
\end{itemize}

\subsection{New Approach to Distributed Stabilization}

Recalling Section~\ref{CSPIS}, the classical separation principle holds because the observer has full access to the input information, which makes the convergence of state estimation errors free from the controller design.
However, this is not the case in a distributed setup. As discussed in Section~\ref{DSPformulation}, each agent typically has access only to some local and incomplete input information.
This is the main reason why separation principle is absent, and why much effort has been devoted to co-designing distributed observers and controllers for stabilization.

In this work, we propose distributed observers where each observer node does not require global input information of the entire system.
Based on such distributed observers, we show that linear state feedback controllers, as well as a class of sliding mode controllers, can be directly employed to achieve distributed stabilization. 
Within this framework, the distributed observers and the state feedback controllers can be designed independently of each other, thereby recovering the benefits of the classical separation principle in a distributed context. 
This framework overcomes the limitations and challenges summarized in Section~\ref{CLACC}. It is also worth noting that the proposed distributed observer design offers a promising approach for achieving cooperative estimation and control in heterogeneous multi-agent systems.

\section{Preliminaries}
\subsection{Notation}
For a vector $x$ and a matrix $X$, $\left\| x \right\|$ and $\left\| X \right\|$ denote the Euclidean norm and the induced 2-norm, respectively.
Let ${\rm{Im}}X$ denote the range or image of $X$.
${\rm{Re}}\lambda (X) < 0$ means that all eigenvalues of $X$ lie in the open left half of the complex plane.
The annihilator of $X$ is a real matrix, whose row vectors are a basis of the left null space of $X$.
$I$ and $0$ denote respectively the identity matrix and zero matrix of appropriate dimensions, whose subscripts are omitted when it causes no ambiguity.
${1_r}$ is a column vector with all $r$ entries equal to one.
For a set of scalars or matrices $\left\{X_i|\ i=1,2,\cdots,N\right\}$, define ${\rm{diag}}({X_i})_{i = 1}^N$ as a matrix formed by arranging the matrices in a block diagonal fashion, and ${\rm{col}}{({X_i})_{i = 1}^N}$ as a matrix formed by stacking them (i.e., 
$\setlength\arraycolsep{1.4pt}
{\left[ {\begin{array}{*{20}{c}}
			{X_1^{\top}}&{X_2^{\top}}& \cdots &{X_N^{\top}}
	\end{array}} \right]^{\top}}$) if the dimensions match. 

\subsection{Problem Formulation}
Consider a linear time-invariant system
\begin{equation}\label{LTIsys}
	\dot x = Ax + Bu + {B_v}v,
\end{equation}
where $x\in \mathbb{R}^n$, $u\in \mathbb{R}^m$, and $v\in \mathbb{R}^{m_{v}}$ are the state, control input, and unknown input vectors, respectively. 
While $v$ is unknown, matrix $B_v$ is known.
This paper is concerned with the following problems.

\begin{problem} \label{PDSE}
	Distributed state estimation
\end{problem}
Consider a group of observer nodes numbered from $1$ to $N$, among which the $i$th observer node has access to local measured output
\begin{equation}\label{localmeasure}
	y_i = C_i x
\end{equation}
with $C_{i}$ having full row rank, and has access to some local control input $u_i$.
Then the input terms in \eqref{LTIsys} can be rewritten as
\begin{equation}\label{splitinput}
	Bu + {B_v}v = B_i u_i + B_{-i}u_{-i}
\end{equation}
with $B_{-i}$ having full column rank and $u_{-i} \in \mathbb{R}^{m_{-i}}$ denoting the input unavailable to the $i$th observer node, as in \cite{G.Yang2022,G.Disaro2025}. 
The communication among observer nodes is mathematically characterized in Section~\ref{commugraph}.
Suppose that the observer nodes exchange state estimates with each other through communication, we aim to design each $i$th observer node ($i=1,\cdots,N$) such that it produces an accurate state estimate $\hat x_i$, i.e., 
\begin{equation*}
	\mathop {\lim }\limits_{t \to \infty } \left\| {{{\hat x}_i}(t) - x(t)} \right\| = 0.
\end{equation*}

\begin{problem} \label{PDOBLFC}
	Distributed observer-based linear feedback control
\end{problem}
In the case where $v(t) \equiv 0$, consider a group of controller nodes indexed from $1$ to $N_c$. The control input term in \eqref{LTIsys} can be written as
\begin{equation}\label{Busplit}
	Bu = \sum\limits_{\iota = 1}^{N_c} {{B^\iota}{u^\iota}},
\end{equation}
where $u^\iota$ denotes the control input vector from the $\iota$th controller node. Given any linear state feedback control law $u^\iota=K^\iota x$ that can stabilize the system \eqref{LTIsys}, we aim to prove that the following control strategy also achieves stabilization:
\begin{equation} \label{linearfeedback}
	u^\iota=K^\iota \hat x_{i^\prime},
\end{equation}
where $\hat x_{i^\prime}$ is the state estimate from any one of the aforementioned observer nodes. 

\begin{problem} \label{PDOBSMC}
	Distributed observer-based sliding mode control
\end{problem}
In the case where $v(t) \ne 0$, consider the following state feedback sliding mode control law that can stabilize system \eqref{LTIsys}: 
\begin{equation}\label{centralsliding}
	u = Kx - \beta h({{B^ \top }Px}),
\end{equation}
where $\beta$ is a scalar, $P$ is a matrix gain, and function ${h}(\cdot)$ is defined as
\begin{equation} \label{hfunc}
	{h}(\omega) = 
	\left\{
	\begin{aligned}
		\left\|\omega\right\|^{-1} \omega,\ \omega &\ne 0 \\
		0,\ \omega &= 0. 
	\end{aligned}
	\right.
\end{equation}
On this basis, we aim to prove that the following control strategy also achieves stabilization:
\begin{equation} \label{slidingfeedback}
	u^\iota = K^\iota \hat x_{i^\prime} - \beta^\iota h\left({(B^\iota)^{\top}}P \hat x_{i^\prime}\right),
\end{equation}
where $\beta^\iota$ is a scalar gain, and $\hat x_{i^\prime}$ is the state estimate from any one of the aforementioned observer nodes.

The separation principle established in this paper can be interpreted as follows: The design of distributed observers is independent of $K^\iota$, $\beta^\iota$, and $P$ in \eqref{linearfeedback} and \eqref{slidingfeedback}. 
Conversely, the distributed controllers can be designed without accounting for the observer dynamics.

\subsection{Communication Links}\label{commugraph}
\begin{figure*}[htpb]\centering
	\centerline{\includegraphics[width=0.75\textwidth]{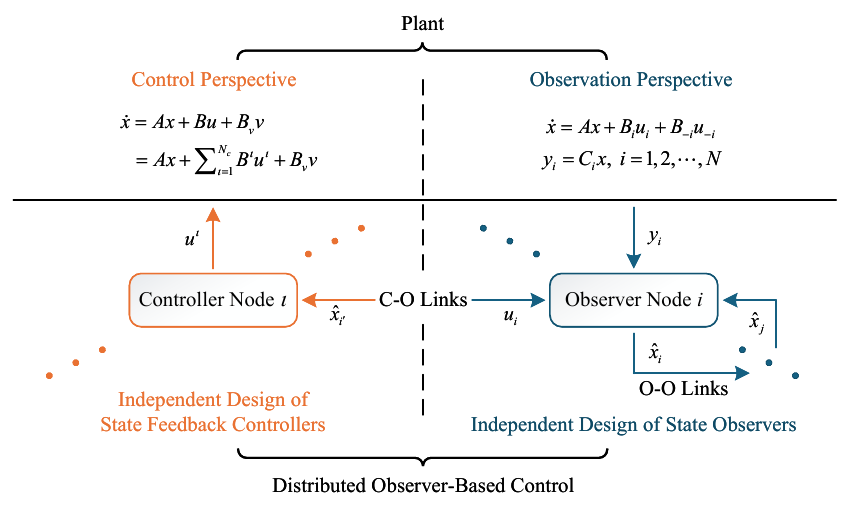}}
	\caption{Overall architecture of the proposed distributed observer-based control framework.}
	\label{ofdobc}
\end{figure*}

As illustrated in Fig.~\ref{ofdobc}, the communication links between controller and observer nodes, which are referred to as C-O links, enable observer nodes to receive control inputs and controller nodes to receive state estimates.
C-O links are responsible for delivering $u_i$ to the $i$th observer node and $\hat x_{i^\prime}$ to the $\iota$th controller node. The topology of C-O links can be designed freely, provided that Assumption~\ref{collecdetec} listed in Section~\ref{mainassums} is satisfied.

The communication links among observer nodes, referred to as O-O links, enable the exchange of state estimates among the observer nodes.
The topology of O-O links can be characterized using an undirected graph introduced below.
This topology can also be designed freely, provided that Assumption~\ref{ooconnected} listed in Section~\ref{mainassums} is satisfied.

A graph $\mathcal{G}=(\mathcal{N},\mathcal{E})$ is composed of a finite nonempty node set $\mathcal{N}=\left\{1,2,\cdots,N\right\}$, and an edge set $\mathcal{E}\subseteq \mathcal{N}\times \mathcal{N}$, whose elements are ordered pairs of nodes. An edge originating from node $j$ and ending at node $i$ is denoted by $(j, i)\in \mathcal{E}$, which represents the direction of the information flow between the two nodes. 
The adjacency matrix of $\mathcal{G}$ is defined as $\mathcal{A}=[a_{ij}]\in \mathbb{R}^{N\times N}$, where $a_{ij}$ is a positive weight of the edge $(j,i)$ when $(j,i)\in \mathcal{E}$, otherwise $a_{ij}$ is zero. 
We assume that the graph has no self-loops, i.e., $a_{ii}=0$, $\forall i \in \mathcal{N}$. The Laplacian matrix $\mathcal{L}=[l_{ij}]\in \mathbb{R}^{N\times N}$ of graph $\mathcal{G}$ is constructed by letting $l_{ii}=\sum\nolimits_{k = 1}^N {{a_{ik}}}$ and $l_{ij}=-a_{ij},\ \forall i,j \in \mathcal{N},\ i\neq j$.
A directed path from node $i$ to node $j$ is a sequence of edges $(i_{k-1},\ i_{k})\in \mathcal{E},\ k=1,2,\cdots,\bar k$, where $i_0=i,\ i_{\bar k}=j$. 
Graph $\mathcal{G}$ is said to be undirected if $a_{ij} = a_{ji}, \forall i,j \in \mathcal{N}$. An undirected graph is said to be connected if there exists at least one directed path from node $i$ to node $j$, $\forall i,j\in\mathcal{N},\ i\neq j$.

\subsection{Supporting Lemmas}
\begin{lemma} \cite{T.Kim2020,G.Cao2023SCL} \label{PosiDef}
	Given $N$ matrices $X_i \in \mathbb{R}^{n\times q_i}$ satisfying $X_i^\mathrm{T}X_i=I_{q_i}$ and a connected undirected graph $\mathcal{G}$, the matrix ${\rm{diag}}^\mathrm{T}(X_i)_{i=1}^N(\mathcal{L}\otimes I_n){\rm{diag}}(X_i)_{i=1}^N$ is positive definite if and only if $\cap_{i=1}^{N} {\rm{Im}}X_i=\left\lbrace 0 \right\rbrace $.
\end{lemma}

\begin{lemma} \label{FUIO}
	Given the matrix triplet $(A,{B_{-i}},{C_i})$ associated with \eqref{LTIsys}, \eqref{localmeasure}, and \eqref{splitinput}, there exists a positive integer $\delta_i $ such that the set $\Omega (\delta_i ) \ne \emptyset $ and $\Omega (\delta_i  + \epsilon ) = \emptyset $ for any positive integer $\epsilon $, where $\Omega (\delta_i)=$\\
	\begin{equation*}
		\left\{\! \! {\begin{array}{*{20}{c}}
				{{T_{id}} \in {\mathbb{R}^{n \times \delta_i }}}\\
				{E_i} \in {\mathbb{R}^{\delta_i  \times \delta_i }}\\
				{F_i} \in {\mathbb{R}^{\delta_i  \times {p_i}}}\\
				{G_i} \in {\mathbb{R}^{\delta_i  \times {p_i}}}
			\end{array} \! \! \left| {\begin{array}{*{20}{c}}
					{T_{id}^{\top}{T_{id}} = {I_{\delta_i} }},\ {G_i}{C_i}{B_{ - i}} = T_{id}^{\top}{B_{ - i}},\\
					{E_i}T_{id}^{\top} + ({G_i}{C_i} - T_{id}^{\top})A\ \ \ \ \ \ \ \ \ \ \ \ \ \\
					\ \ \ \ \ \ \ \ \ \ \ \ \ \ + ({F_i} - {E_i}{G_i}){C_i} = 0,\\
					{{\rm{Re}}\lambda ({E_i}) < 0{\rm{\ or \ }}{E_i} = 0_{p_i \times p_i}}
			\end{array}} \right.} \! \! \right\}.
	\end{equation*}
	A matrix quadruplet $\left( {{T_{id}},{E_i},{F_i},{G_i}} \right) \in \Omega (\delta_i )$ can be computed using the algorithm provided in Section~\ref{appendix-FUIO}.
\end{lemma}
\begin{rem}
	Lemma~\ref{FUIO} follows the main results in \cite{S.Sundaram2008}. It is guaranteed that there exists no matrix quadruple $\left( {{T_{id}},{E_i},{F_i},{G_i}} \right) \in \Omega$, of which the rank of $T_{id}$ is higher than $\delta_i$. This indicates that $T_{id}^\top x \in \mathbb{R}^{\delta_i}$ captures the maximum amount of state information that can be reconstructed (see Table~\ref{comparetable} for details).  
\end{rem}

\begin{lemma}\cite{H.K.Khalil2002}\label{barbalaslemma}
	Let $f: \mathbb{R} \to \mathbb{R}$ be a uniformly continuous function on $[0,\infty)$. Suppose that ${\lim _{t \to \infty }}\int_0^t {f(\tau ){\rm{d}}\tau } $ exists and is finite, then ${\lim _{t \to \infty }} f(t) = 0 $.
\end{lemma}

\begin{lemma}\cite{H.K.Khalil2002}\label{comparelemma}
Consider the scalar differential equation
$$\dot x_a(t) = f\left(x_a(t)\right),\ x_a(t_0)=x_{a0},$$
where $f(x_a)$ is locally Lipschitz for all $x_a \in \mathbb{R}$. Let $x_b$ be a continuous function whose derivative satisfies
$$\dot x_b(t) \le f\left( x_b(t)\right),\ x_b(t_0) \le x_{a0}.$$
Then $x_a(t) \le x_b(t)$, for all $t \ge t_0$.
\end{lemma}

\subsection{Main Assumptions} \label{mainassums}
\begin{assu}\label{vbound}
	The unknown input $v(t)$ is bounded, i.e., $\mathop {\max }\limits_t \left\| v (t) \right\| \le \bar v$.
\end{assu} 

\begin{assu}\label{ubound}
	For each observer node $i$, the portion of the input unavailable to that node is bounded, i.e., $\mathop {\max }\limits_t \left\| u_{-i} (t) \right\| \le \bar u_{-i},\ \forall i \in \left\{1,2,\cdots,N\right\}$.
\end{assu}

\begin{assu}\label{controllable}
	The pair $(A,B)$ is stabilizable.
\end{assu}

\begin{assu}\label{matchcondition}
	The channels of the control input and unknown input are matched; that is, there exists a matrix $X_v$ such that $B X_v=B_v$.
\end{assu}

\begin{assu}\label{ooconnected}
	The observer nodes communicate according to a connected graph $\mathcal{G}$.
\end{assu}

\begin{assu}\label{collecdetec}
	The matrix triplet $(A,{B_{-i}},{C_i})$ is collectively strongly detectable, i.e., $\sum\nolimits_{i = 1}^N {{\mathop{\rm Im}\nolimits} {T_{id}}}  = \mathbb{R}^n$.
\end{assu}

\begin{rem}
	Table~\ref{comparetable} clarifies that Assumption~\ref{collecdetec} serves as a natural extension of the classical detectability condition in the presence of unavailable inputs. If $B_{-i} = 0$ for all $i \in \mathcal{N}$, then Assumption~\ref{collecdetec} reduces to the requirement that the pair $\left( {A,{\rm{col}}({C_i})_{i = 1}^N} \right)$ is detectable. When there exists at least one $i \in \mathcal{N}$ for which $B_{-i} \neq 0$, a sufficient (yet not necessary) condition for Assumption~\ref{collecdetec} is that ${\rm{col}}({C_{i}})_{i = 1}^N$ has full column rank. In addition, Section~\ref{AIHMAS} illustrates an application scenario in which Assumption~\ref{collecdetec} always holds.
\end{rem}

\begin{table*}
	\centering
\caption{Comparison between detectable and strongly detectable subspaces.}
\label{comparetable}
\renewcommand{\arraystretch}{2}
\begin{tabular}{c|c|c}
	\hline
	System & \small$\left\{ \begin{aligned}
		\dot{x} &= A x \\[-2pt]
		y_i &=C_i x 
	\end{aligned}\right.$ & \small$\left\{ \begin{aligned}
		\dot{x} &= A x + B_{-i}u_{-i}\\[-2pt]
		y_i &=C_i x 
	\end{aligned}\right.$\\
	\hline
	Decomposition& Detectability decomposition & Strong detectability decomposition, i.e., Lemma~\ref{FUIO} \\
	\hline
	Result& Detectable subspace ${\rm Im}T_{id}$ & Strongly detectable subspace ${\rm Im}T_{id}$\\
	\hline
	Meaning & Functional $T_{id}^\top x$ can be reconstructed from output $y_i$ &Functional $T_{id}^\top x$ can be reconstructed from output $y_i$\\
	\hline
	Assumption & $\sum\nolimits_{i = 1}^N {{\mathop{\rm Im}\nolimits} {T_{id}}}  = \mathbb{R}^n$ $ \Leftrightarrow$ $\left( {A,{\rm{col}}({C_i})_{i = 1}^N} \right)$ is detectable&$\sum\nolimits_{i = 1}^N {{\mathop{\rm Im}\nolimits} {T_{id}}}  = \mathbb{R}^n$ $\Leftrightarrow$ $\left(A,{B_{-i}},{C_i}\right)$ is collectively strongly detectable\\
	\hline
	Connection & \multicolumn{ 2 }{c}{$\left( {A,{\rm{col}}({C_i})_{i = 1}^N} \right)$ is detectable $\Leftrightarrow$ $\left(A,0,{C_i}\right)$ is collectively strongly detectable}\\
	\hline
\end{tabular}
\end{table*}

Not all assumptions are required in every section of the paper. In particular:
\begin{itemize}
	\item In Section~\ref{SFDOWUGI}, Problem~\ref{PDSE} is addressed under Assumptions~\ref{vbound}, \ref{ubound}, \ref{ooconnected}, and \ref{collecdetec}, without requiring prior knowledge of $\bar v$ or $\bar u_{-i}$.
	\item In Section~\ref{SDOBLFC}, Problem~\ref{PDOBLFC} is addressed under Assumptions~\ref{controllable}, \ref{ooconnected}, and \ref{collecdetec}.
	\item In Section~\ref{SDOBSMC}, Problem~\ref{PDOBSMC} is addressed under Assumptions~\ref{vbound}, \ref{controllable}, \ref{matchcondition}, \ref{ooconnected}, and \ref{collecdetec}, with or without prior knowledge of $\bar v$.
\end{itemize}

\section{Fully Distributed State Estimation Without Using Global Inputs}\label{SFDOWUGI}
\subsection{Theoretical Results}
This section develops fully distributed observers, where each observer node can provide an estimate of the full state of system \eqref{LTIsys}.
The term ``fully'' indicates that the design of each observer node only relies on locally available information, including local input, local output, and state estimates received from its neighbors. 
The $i$th observer node is designed as
\begin{subequations}\label{obidynamic}
	\begin{align}
		{{\dot z}_i} &= {{\bar E}_i}{z_i} + {{\bar F}_i}{y_i} + {{\bar B}_i}{u_i} - {H_i}\left( {\sum\limits_{j = 1}^N {{a_{ij}}({{\hat x}_i} - {{\hat x}_j})} } \right) \label{zidynamic}\\
		{{\hat x}_i} &= {z_i} + {{\bar G}_i}{y_i}, \label{hatxieq}
	\end{align}
\end{subequations}
where $z_i$ with initial value $z_i(0) = 0$ is the state of the node dynamics, and ${\hat x}_i$ is the state estimate of $x$ produced by the $i$th observer node. 
The matrix gains in \eqref{obidynamic} are designed as
\begin{subequations}\label{barEFGB}
\begin{align}
	{{\bar E}_i} &= {T_{id}}{E_i}T_{id}^{\top} + {T_{iu}}T_{iu}^{\top}A \label{barE_i}\\
	{{\bar F}_i} &= {T_{id}}{F_i} + {T_{iu}}T_{iu}^{\top}A{{\bar G}_i} \label{barF_i}\\
	{{\bar G}_i} &= {T_{id}}{G_i} \label{barG_i}\\
	{{\bar B}_i} &= (I - {\bar G}_i{C_i}){B_i}, \label{barB_i}
\end{align}
\end{subequations}
where $T_{id}$, $E_i$, $F_i$ and $G_i$ are obtained from Lemma~\ref{FUIO}, and $T_{iu}$ is a matrix satisfying 
\begin{equation} \label{TidTiu}
	 T_{id}^\top T_{iu} = 0,\ T_{iu}^\top T_{iu} = I_{n-\delta_i},\ \text{and}\ T_{iu} T_{iu}^\top = I_n - T_{id} T_{id}^\top.
\end{equation}
For notational simplicity, let $\varepsilon_{iu} =T_{iu}^\top \sum\nolimits_{j = 1}^N {a_{ij} (\hat x_i - \hat x_j)}$.
Based on ${h}(\cdot)$ defined in \eqref{hfunc}, function $H_i(\cdot)$ in \eqref{zidynamic} is designed as
\begin{equation} \label{H_idefine}
	{H_i}(\cdot) = {\gamma_i}{T_{iu}}\varepsilon_{iu} + {\gamma _{is}}{T_{iu}}{h}(\varepsilon_{iu}),
\end{equation}
where $\gamma_i$ and $\gamma_{is}$ are scalar gains evolving according to the following adaptive laws:
\begin{subequations}\label{gammaiis}
\begin{align}
	{{\dot \gamma }_i} &= {\phi _i}{\left\| \varepsilon_{iu} \right\|^2} \\
	{{\dot \gamma }_{is}} &= {\phi _{is}}\left\| \varepsilon_{iu} \right\| 
\end{align}
\end{subequations} 
with step sizes ${\phi _{i}}$, ${\phi _{is}}$ and initial values $\gamma_i(0)$, $\gamma_{is}(0)$ chosen as positive real numbers.

\begin{thm} \label{observerthm}
	Under Assumptions \ref{vbound}, \ref{ubound}, \ref{ooconnected}, and \ref{collecdetec}, the distributed observers, with each observer node governed by the node dynamics in \eqref{obidynamic}, can produce accurate state estimates for system \eqref{LTIsys}, i.e.,
	\begin{equation*}
		\mathop {\lim }\limits_{t \to \infty } \left\| {{{\hat x}_i}(t) - x(t)} \right\| = 0,\ \forall i \in \mathcal{N}.
	\end{equation*}
	Moreover, adaptive gains $\gamma_i$ and $\gamma_{is}$ remain bounded, $\forall i \in \mathcal{N}$.
\end{thm}
See Section~\ref{proofobserverthm} for the proof of Theorem~\ref{observerthm}.

\subsection{Application in Heterogeneous Multi-Agent Systems}\label{AIHMAS}

Consider a group of $N$ agents that have different general linear dynamics. The dynamics of the $i$th agent are described by
\begin{subequations}\label{HMAS}
\begin{align}
	\dot{\breve{x}}_i =& \breve{A}_i \breve{x}_i + \breve{B}_i u_i \\
	{y}_i =& \breve{C}_i \breve{x}_i,\ i \in \mathcal{N},
\end{align}
\end{subequations}
where $\breve{x}_i \in \mathbb{R}^{n_i}$ is the state, $u_i \in \mathbb{R}^{m_i}$ is the control input, and $y_i \in \mathbb{R}^{p_i}$ is the measured output. The dynamics of the overall multi-agent system take the form of \eqref{LTIsys}, \eqref{localmeasure}, and \eqref{splitinput}, where $x = {\rm{col}}(\breve{x}_i)_{i = 1}^N$, $A = {\rm{diag}}(\breve{A}_i)_{i = 1}^N$, $B = {\rm{diag}}(\breve{B}_i)_{i = 1}^N$,
\begin{align*}
	{B_i} =& {\left[ {\begin{array}{*{20}{c}}
				{{0_{{m_i} \times \sum\nolimits_{q = 1}^{i - 1} {{n_q}} }}}&{\breve{B}_i^{\top}}&{{0_{{m_i} \times \sum\nolimits_{q = i + 1}^N {{n_q}} }}}
		\end{array}} \right]^{\top}} \\
	C_i =& \left[ {\begin{array}{*{20}{c}}
			{{0_{{p_i} \times \sum\nolimits_{q = 1}^{i - 1} {{n_q}} }}}&{\breve{C}_i}&{{0_{{p_i} \times \sum\nolimits_{q = i + 1}^N {{n_q}} }}}
	\end{array}} \right].
\end{align*}
Suppose the $i$th agent only has access to its input $u_i$ and output $y_i$.
If $(\breve{A}_i, \breve{C}_i)$ is detectable for all $i \in \mathcal{N}$, then Assumption~\ref{collecdetec} is satisfied, with
\begin{equation*}
	{T_{id}} = {\left[ {\begin{array}{*{20}{c}}
				{{0_{{n_i} \times \sum\nolimits_{q = 1}^{i - 1} {{n_q}} }}}&{{I_{{n_i}}}}&{{0_{{n_i} \times \sum\nolimits_{q = i + 1}^N {{n_q}} }}}
		\end{array}} \right]^{\top}}.
\end{equation*}
Under this condition, observer node dynamics \eqref{obidynamic} reduce to the following form:
\begin{equation}\label{obidynamicMAS}
	{{\dot {\hat x}}_i} = {{\bar E}_i}{{\hat x}_i} + {{\bar F}_i}{y_i} + {B_i}{u_i} - {H_i}\left( {\sum\limits_{j = 1}^N {{a_{ij}}({{\hat x}_i} - {{\hat x}_j})} } \right) 
\end{equation}
with initial value ${\hat x}_i(0) = 0$, where function $H_i(\cdot)$ is the same as in \eqref{H_idefine} and \eqref{gammaiis},
\begin{align*}
	{\bar E}_i =& T_{id} (\breve{A}_i + \breve{L}_i \breve{C}_i)T_{id}^\top + T_{iu} T_{iu}^\top A,\ {{\bar F}_i} = -{T_{id}}{\breve{L}_i},
\end{align*} 
and $\breve{L}_i$ is chosen such that $\breve{A}_i + \breve{L}_i \breve{C}_i$ has eigenvalues
with negative real parts. Based on Theorem~\ref{observerthm}, the following theorem directly follows.

\begin{thm} \label{HMASthm}
	Consider a heterogeneous multi-agent system \eqref{HMAS}, where $(\breve{A}_i, \breve{C}_i)$ is detectable and $u_i$ is bounded, $\forall i \in \mathcal{N}$. If the $i$th agent implements observer dynamics \eqref{obidynamicMAS}, then it can produce an accurate state estimate for multi-agent system \eqref{HMAS}, i.e.,
	\begin{equation*}
		\mathop {\lim }\limits_{t \to \infty } \left\| {{{\hat x}_i}(t) - x(t)} \right\| = 0,\ \forall i \in \mathcal{N}.
	\end{equation*}
	Moreover, adaptive gains $\gamma_i$ and $\gamma_{is}$ remain bounded, $\forall i \in \mathcal{N}$.	
\end{thm}

This result implies that, in a heterogeneous multi-agent system, each agent is able to reconstruct not only its own state but also the states of all other agents, without requiring access to their control inputs or output measurements.
Based on others' real-time state information, each agent is omniscient and so can autonomously plan its own actions.
Such capability can help a group of agents engage in collaborative tasks that are more complex than consensus-seeking.

\subsection{Numerical Example}\label{NE1}
\begin{figure}[htpb]\centering
	\centerline{\includegraphics[width=0.4\textwidth]{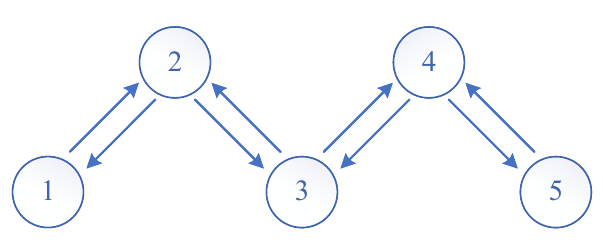}}
	\caption{O-O Links in Section~\ref{NE1}.}
	\label{MASGraph}
\end{figure}

\begin{figure}[htpb]\centering
	\centerline{\includegraphics[width=0.45\textwidth]{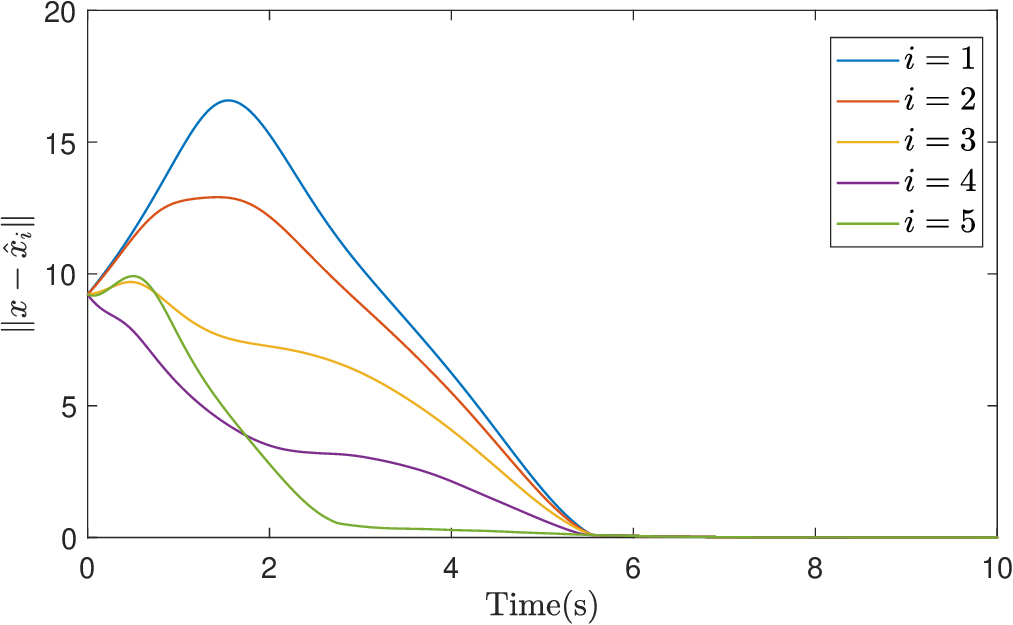}}
	\caption{State estimation error norms in Section~\ref{NE1}.}
	\label{MASError}
\end{figure}

\begin{figure}[htpb]\centering
	\centerline{\includegraphics[width=0.45\textwidth]{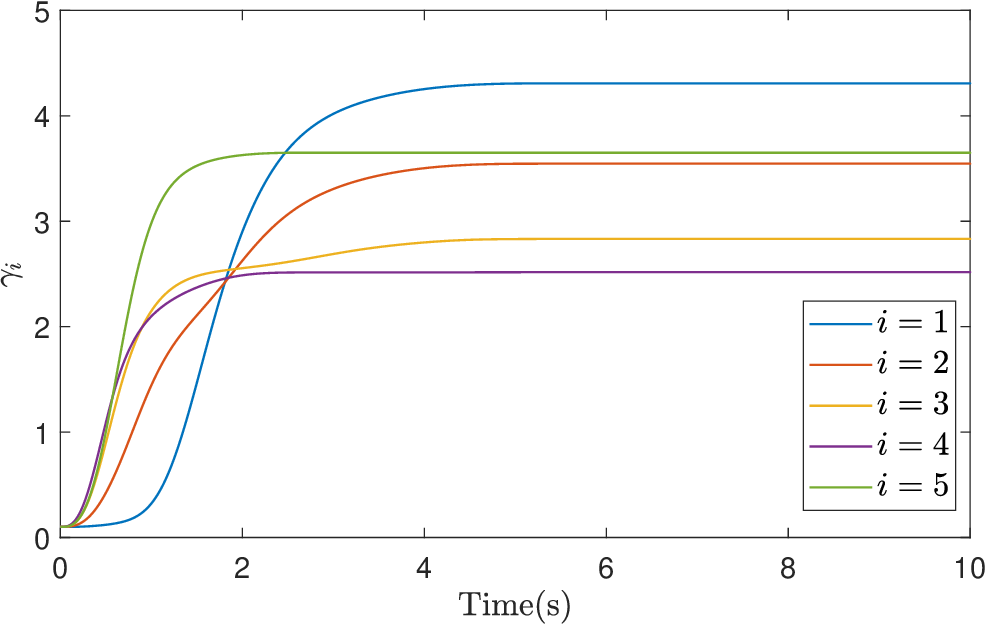}}
	\caption{Adaptive gains in the distributed observers in Section~\ref{NE1}.}
	\label{MASgamma}
\end{figure}

\begin{figure}[htpb]\centering
	\centerline{\includegraphics[width=0.45\textwidth]{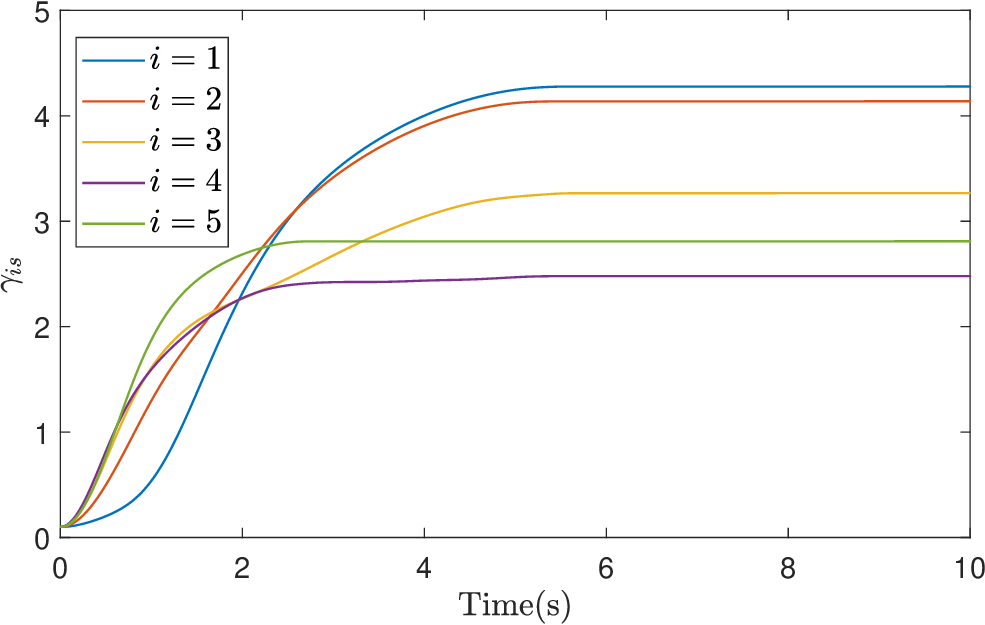}}
	\caption{Adaptive gains in the distributed observers in Section~\ref{NE1}.}
	\label{MASgammas}
\end{figure}

Consider a heterogeneous multi-agent system \eqref{HMAS} composed of five agents, whose system matrices are
\begin{align*}
	\breve{A}_1 =& \breve{A}_2 = 0,\ \breve{B}_1 = \breve{B}_2 = 1,\ \breve{C}_1 = \breve{C}_2 = 1\\
	\breve{A}_3 =& \breve{A}_4 = \begin{bmatrix} 
		0 & 1 \\ 
		0 & 0   
	\end{bmatrix},\ \breve{B}_3 = \breve{B}_4 = \begin{bmatrix} 
	0 \\ 
	1 
	\end{bmatrix},\ \breve{C}_3^\top = \breve{C}_4^\top = \begin{bmatrix} 
	1 \\ 
	0 
	\end{bmatrix}\\
	\breve{A}_5 =& \begin{bmatrix} 
		0 & 1 & 0\\ 
		0 & 0 & 1\\
		0 & 0 & 0 
	\end{bmatrix},\ \breve{B}_5 = \begin{bmatrix} 
	0 \\ 
	0 \\
	1
	\end{bmatrix},\ \breve{C}_5^\top = \begin{bmatrix} 
	1 \\ 
	0 \\
	0
	\end{bmatrix}.
\end{align*} 
Set the initial states of the agents as 
\begin{align*}
	\breve{x}_1(0) =& -1,\ \breve{x}_2(0) = -2,\ 
	\breve{x}_3(0) = \begin{bmatrix} 
		-3&-4 \\ 
	\end{bmatrix}^\top\\ 
	\breve{x}_4(0) =& \begin{bmatrix} 
	5&4 \\ 
	\end{bmatrix}^\top,\ \breve{x}_5(0) = \begin{bmatrix} 
	3&2&1 \\ 
	\end{bmatrix}^\top.
\end{align*}
Set the control inputs as
\begin{equation*}
	u_i = 0.5(i-1)\sin\left[(6-i)t\right],\ i=1,2,3,4,5.
\end{equation*}
In observer node dynamics \eqref{obidynamicMAS}, design gain $\breve{L}_i$ as
\begin{equation*}
	\breve{L}_i = - \breve{X}_i \breve{C}_i^\top,\ i=1,2,3,4,5,
\end{equation*}
where $\breve{X}_i$ is the unique solution of algebraic Riccati equation
\begin{equation*}
	(\breve{A}_i + 0.2 I)\breve{X}_i + \breve{X}_i(\breve{A}_i + 0.2 I)^\top - \breve{X}_i \breve{C}_i^\top \breve{C}_i \breve{X}_i + I =0.
\end{equation*}
For adaptive gains $\gamma_i$ and $\gamma_{is}$, set their initial values as 
\begin{align*}
	\gamma_i(0) = \gamma_{is}(0) = 0.1
\end{align*}
and update step sizes as 
\begin{align*}
	{\phi _{i}} = 0.2,\ {\phi _{is}} = 0.5,\ i=1,2,3,4,5.
\end{align*}
Let the agents communicate with each other according to the graph shown in Fig.~\ref{MASGraph}, in which the weights of edges are all set as $1$.
The simulation results in Fig.~\ref{MASError} show that each agent is able to correctly estimate the states of all agents. Results in Fig.~\ref{MASgamma} and Fig.~\ref{MASgammas} show that adaptive gains are bounded all the time.

\section{Distributed Observer-Based Linear Feedback Control}\label{SDOBLFC}
\subsection{Theoretical Results}
The distributed observers proposed in Section~\ref{SFDOWUGI} are guaranteed to be effective, on the premise that the control inputs remain bounded.
However, when state feedback controllers are implemented by replacing the true state with the estimated state generated by the distributed observers, it is not immediately clear whether the resulting control inputs will remain bounded.
Therefore, the convergence of state estimation errors cannot be deduced directly from Theorem~\ref{observerthm}. 
To obtain rigorous guarantees, it is necessary to analyze the closed-loop system composed of the plant, the distributed observers, and the distributed controllers.
\begin{thm} \label{lfthm}
	Consider the case where $v(t)\equiv0$, if Assumptions~\ref{controllable}, \ref{ooconnected}, and \ref{collecdetec} hold, and if a linear state feedback control law 
	\begin{equation*}
		u^\iota=K^\iota x,\ \iota=1,2,\cdots,N_c
	\end{equation*}
	taken by the controller nodes can stabilize system \eqref{LTIsys}, i.e., achieve
	\begin{equation*}
		\mathop {\lim }\limits_{t \to \infty } \left\| {x(t)} \right\| = 0,
	\end{equation*}
	then the distributed observer-based control law
	\begin{equation*}
		u^\iota = K^\iota \hat x_{i^\prime},\ \iota=1,2,\cdots,N_c,
	\end{equation*}
	where $\hat x_{i^\prime}$ is the state estimate generated by any observer node designed in Section~\ref{SFDOWUGI}, also stabilizes system~\eqref{LTIsys}. Moreover, the state estimation error $e_i$ converges to zero, and adaptive gains $\gamma_i$ and $\gamma_{is}$ remain bounded, $\forall i \in \mathcal{N}$.  
\end{thm}
See Section~\ref{prooflfthm} for the proof of Theorem~\ref{lfthm}.

In the proof of Theorem~\ref{lfthm}, it appears that the value of $\gamma^*_s$ (whether $\gamma^*_s = 0$ or $\gamma^*_s > 0$) does not affect the validity of the result.
In fact, Theorem~\ref{lfthm} still holds even if $\gamma_{is}$ is designed as $\gamma_{is}(0)= 0$ and $\dot \gamma_{is} = 0$ for all $i \in \mathcal{N}$.
This observation motivates a detailed examination of the role of the discontinuous term in \eqref{H_idefine} during stabilization.
From the proof, it can be noticed that the state estimation errors and the system state can only converge to zero simultaneously.
This implies that the distributed observers are unable to provide meaningful information about the true state until the true state itself approaches zero.

However, the behavior becomes different once the discontinuous term is incorporated.
Since the Lyapunov function \eqref{V_x} is shown to be bounded, it is guaranteed that the control inputs remain bounded. 
By Theorem~\ref{observerthm}, this boundedness ensures that the decay of the state estimation errors does not depend on whether the control input or the true state tends to zero.

Moreover, the discontinuous term will immediately become indispensable, when the tracking problem is considered instead of the stabilization problem. To illustrate this point, let system \eqref{LTIsys} track the following reference system
\begin{equation} \label{refxdy}
	\dot x_r = Ax_r + \sum\limits_{\iota  = 1}^{{N_c}} {{B^\iota } \left({K^\iota }x_r + r^\iota\right) },
\end{equation}
where $K^\iota$ is a gain that can stabilize system \eqref{LTIsys}, and $r^\iota$ is a bounded time-varying reference signal vector.
\begin{thm} \label{reftrackthm}
	Consider the case where $v(t)\equiv0$, and implement the distributed observer-based control law
	\begin{equation*}
		u^\iota = K^\iota \hat x_{i^\prime} + r^\iota,\ \iota=1,2,\cdots,N_c,
	\end{equation*}
	where $\hat x_{i^\prime}$ is the state estimate generated by any observer node designed in Section~\ref{SFDOWUGI}.
	If Assumptions~\ref{controllable}, \ref{ooconnected}, and \ref{collecdetec} hold, then the state tracking error $e_r:=x-x_r$ converges to zero. Moreover, the state estimation error $e_i$ converges to zero, and adaptive gains $\gamma_i$ and $\gamma_{is}$ remain bounded, $\forall i \in \mathcal{N}$. 
\end{thm}

The proof of Theorem~\ref{reftrackthm} is omitted due to its similarity to the proof of Theorem~\ref{lfthm}. 

In the above tracking problem, the control input will not converge to zero, unless the reference signal itself converges to zero.
Consequently, the results in Theorem~\ref{reftrackthm} cannot be achieved without the discontinuous term in \eqref{H_idefine}.
This is illustrated by the simulation results in what follows.

\subsection{Numerical Example} \label{NE2}
\begin{figure}[htpb]\centering
	\centerline{\includegraphics[width=0.4\textwidth]{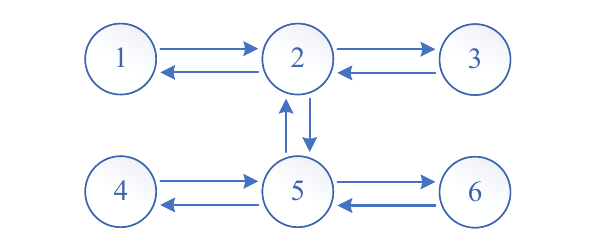}}
	\caption{O-O links in Section~\ref{NE2} and \ref{NE3}.}
	\label{LFCGraph}
\end{figure}

\begin{figure}[htpb]\centering
	\centerline{\includegraphics[width=0.45\textwidth]{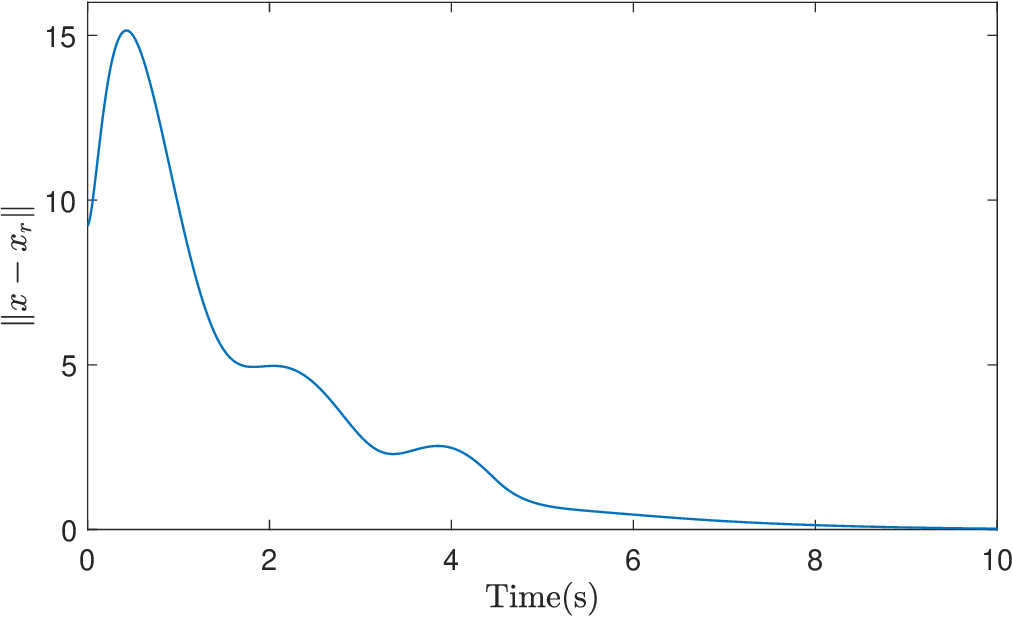}}
	\caption{Tracking error norm in Section~\ref{NE2}.}
	\label{LFCTrackingError}
\end{figure}

\begin{figure}[htpb]\centering
	\centerline{\includegraphics[width=0.45\textwidth]{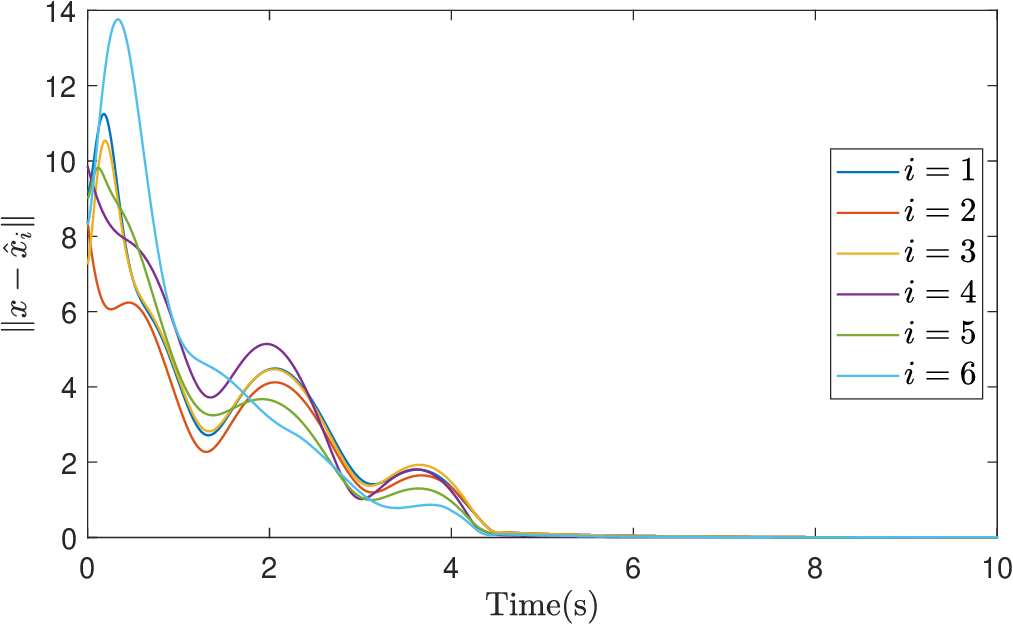}}
	\caption{State estimation error norms in Section~\ref{NE2}.}
	\label{LFCEstimationError}
\end{figure}

\begin{figure}[htpb]\centering
	\centerline{\includegraphics[width=0.45\textwidth]{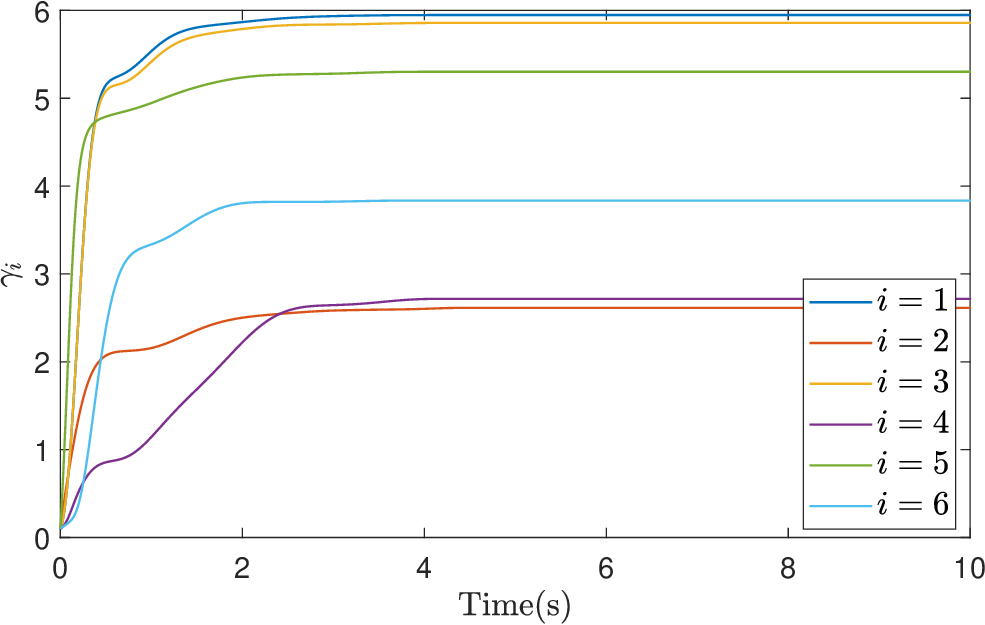}}
	\caption{Adaptive gains in the distributed observers in Section~\ref{NE2}.}
	\label{LFCgamma}
\end{figure}

\begin{figure}[htpb]\centering
	\centerline{\includegraphics[width=0.45\textwidth]{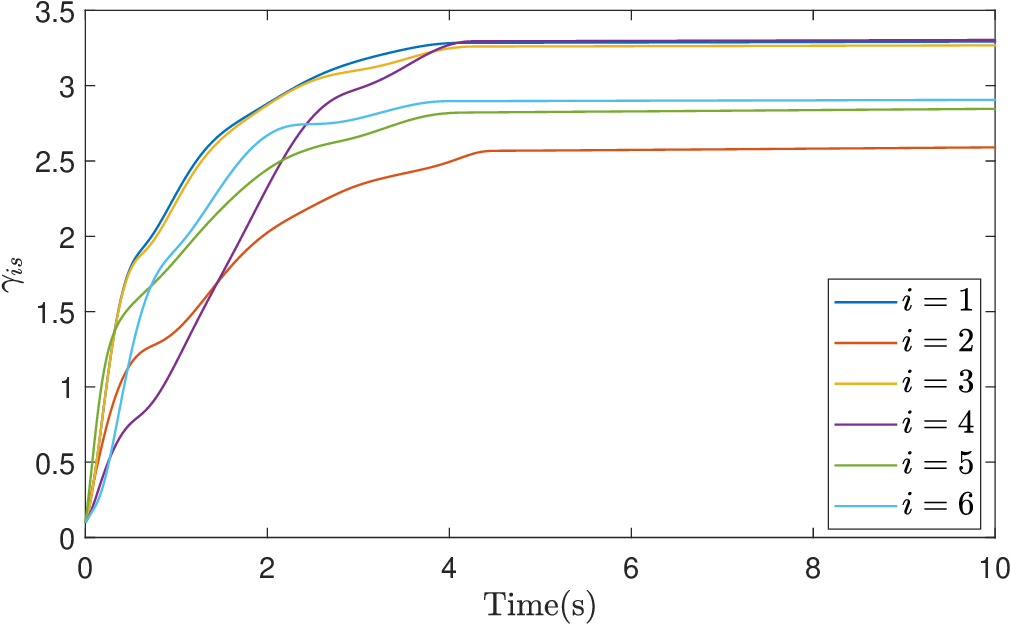}}
	\caption{Adaptive gains in the distributed observers in Section~\ref{NE2}.}
	\label{LFCgammas}
\end{figure}

Consider the tracking problem for system \eqref{LTIsys} in the case where $v(t)\equiv0$. Five controller nodes and six observer nodes are involved. The parameters of system \eqref{LTIsys} and reference system \eqref{refxdy} are chosen as
\begin{align*}
	A =& 
	\begin{bmatrix} 
		0&1&0&1&0&0&0&0&0 \\ 
		0&0&0&0&0&0&0&0&1 \\
		0&0&0&1&0&0&0&0&0 \\
		0&0&0&0&0&0&1&0&0 \\
		0&0&0&0&0&1&0&0&0 \\
		0&0&0&0&0&0&0&0&0 \\
		0&0&0&0&0&0&0&1&0 \\
		0&0&0&0&0&0&0&0&1 \\
		0&0&0&0&0&0&0&0&0 \\
	\end{bmatrix} \\ 
	B^1 =& 
	\begin{bmatrix} 
		0&1&0&0&0&0&0&0&0
	\end{bmatrix}^\top \\ 
	B^2 =& 
	\begin{bmatrix} 
		0&0&0&1&0&0&0&0&0
	\end{bmatrix}^\top \\ 
	B^3 =& 
	\begin{bmatrix} 
		1&0&0&0&1&1&0&0&0
	\end{bmatrix}^\top \\
	B^4 =& 
	\begin{bmatrix} 
		0&0&1&0&0&0&0&0&1
	\end{bmatrix}^\top \\
	B^5 =& 
	\begin{bmatrix} 
		1&0&0&0&0&0&0&0&0
	\end{bmatrix}^\top \\
	x(0) =&
	\begin{bmatrix} 
		-1&-2&-3&-4&5&4&3&2&1
	\end{bmatrix}^\top\\
	x_r (0) =&
	\begin{bmatrix} 
		0&0&0&0&0&0&0&0&0
	\end{bmatrix}^\top\\
	K^\iota =& -(B^\iota)^\top X,\ r^\iota = \sin(t+\iota),\ \iota = 1,2,3,4,5,
\end{align*}
where $X$ is the unique solution of algebraic Riccati equation
\begin{equation*}
	(A + 0.2 I)^\top X + X(A + 0.2 I) - X B B^\top X + I =0,
\end{equation*}
with $B=\begin{bmatrix} 
	B^1&B^2&\cdots&B^5
\end{bmatrix}$. 
The $i$th observer node receives the local output $y_i$ in \eqref{localmeasure}, with 
\begin{align*}
	C_1 =& 
	\begin{bmatrix} 
		1&0&0&0&0&0&0&0&0
	\end{bmatrix} \\ 
	C_2 =& 
	\begin{bmatrix} 
		0&0&1&0&0&0&0&0&0
	\end{bmatrix} \\ 
	C_3 =& 
	\begin{bmatrix} 
		0&0&0&0&1&0&0&0&0
	\end{bmatrix} \\
	C_4 =& 
	\begin{bmatrix} 
		0&0&0&0&0&0&1&0&0
	\end{bmatrix} \\
	C_5 =& 
	\begin{bmatrix} 
		0&1&0&0&0&0&0&0&0
	\end{bmatrix} \\
	C_6 =& 
	\begin{bmatrix} 
		0&0&0&1&0&0&0&0&0
	\end{bmatrix}.
\end{align*} 
The local control inputs available to observer nodes are
\begin{equation*}
	u_i = u^i = K^i \hat x_{i} + r^i\ (i=1,2,3,4,5),\ u_6 = K^5 \hat x_{5} + r^5,
\end{equation*}
whose associated input channels are
\begin{equation*}
	B_i = B^i\ (i=1,2,3,4,5),\ B_6 = B^5.
\end{equation*}
Accordingly, $B_{-i}$ can be determined. Based on the algorithm provided in Section~\ref{appendix-FUIO}, it can be obtained that 
\begin{align*}
	T_{1d} =& 
	\begin{bmatrix} 
		1&0&0&0&0&0&0&0&0
	\end{bmatrix}^\top \\
	T_{2d} =& 
	\begin{bmatrix} 
		0&0&0&0&0&0&0.5&-0.7071&0.5\\
		0&0&0&-0.7071&0&0&0.5&0&-0.5\\
		0&0&-1&0&0&0&0&0&0
	\end{bmatrix}^\top \\
	T_{3d} =& 
	\begin{bmatrix} 
		0&0&0&0&0&-1&0&0&0\\
		0&0&0&0&1&0&0&0&0
	\end{bmatrix}^\top \\
	T_{4d} =& 
	\begin{bmatrix} 
		0&0&0&0&0&0&0&1&0\\
		0&0&0&0&0&0&0&0&1\\
		0&0&0&0&0&0&-1&0&0
	\end{bmatrix}^\top \\
	T_{5d} =& 
	\begin{bmatrix} 
		0&1&0&0&0&0&0&0&0
	\end{bmatrix}^\top \\
	T_{6d} =& 
	\begin{bmatrix} 
		0&0&0&1&0&0&0&0&0
	\end{bmatrix}^\top.
\end{align*}
For brevity, the values of matrices $E_i$, $F_i$, $G_i$ are not listed here.
The adaptive gains $\gamma_i$ and $\gamma_{is}$ are initialized as:
\begin{align*}
	\gamma_i(0) = \gamma_{is}(0) = 0.1
\end{align*}
and their update step sizes are chosen as 
\begin{align*}
	{\phi _{i}} = 0.2,\ {\phi _{is}} = 0.5,\ i=1,2,3,4,5,6.
\end{align*}
The observer nodes communicate according to the graph shown in Fig.~\ref{LFCGraph}, in which the weights of edges are all set as $1$.
The simulation results in Fig.~\ref{LFCTrackingError} and Fig.~\ref{LFCEstimationError} show that the tracking error and state estimation errors converge to zero. 
The adaptive gains remain bounded throughout the simulation, as shown in Fig.~\ref{LFCgamma} and Fig.~\ref{LFCgammas}.

\begin{figure}[htpb]\centering
	\centerline{\includegraphics[width=0.45\textwidth]{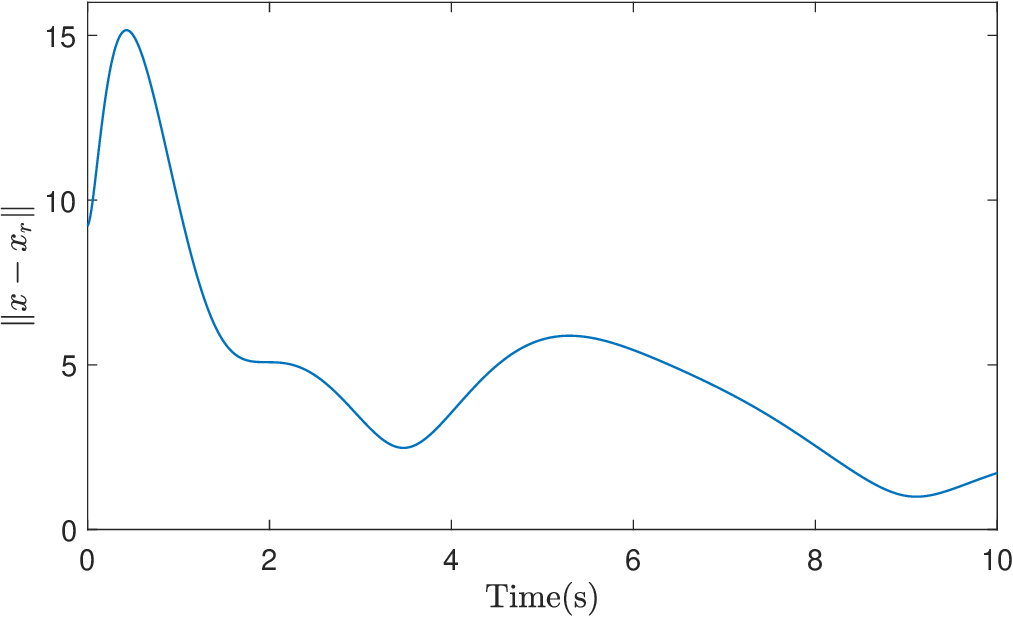}}
	\caption{Tracking error norm in case $\gamma_{is} \equiv 0$ in Section~\ref{NE2}.}
	\label{LFCTrackingError-Coma}
\end{figure}

\begin{figure}[htpb]\centering
	\centerline{\includegraphics[width=0.45\textwidth]{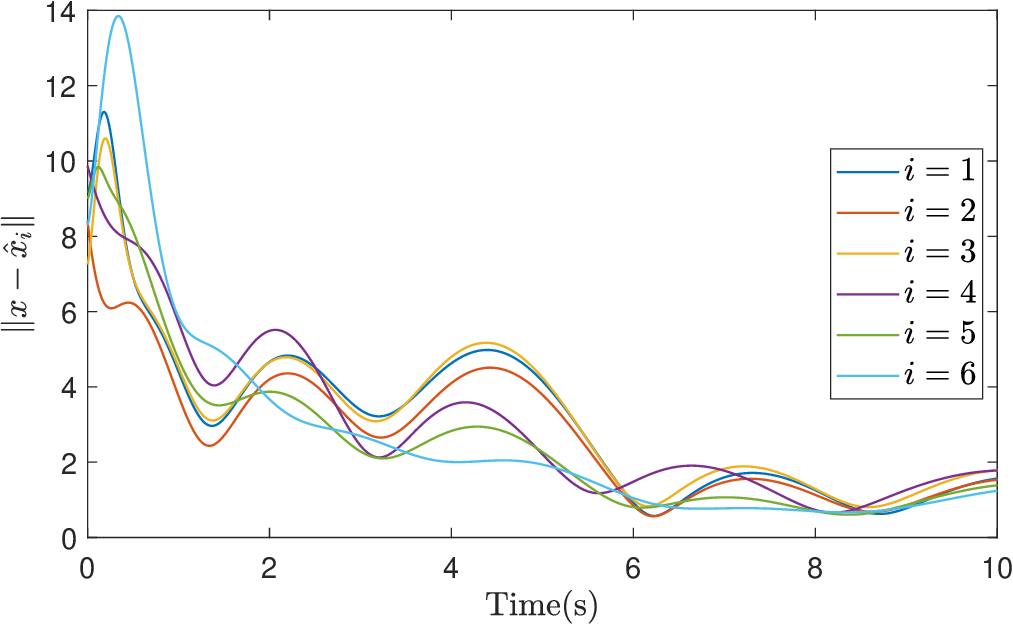}}
	\caption{State estimation error norms in case $\gamma_{is} \equiv 0$ in Section~\ref{NE2}.}
	\label{LFCEstimationError-Coma}
\end{figure}

To demonstrate the indispensability of the discontinuous term in \eqref{H_idefine}, we set $\gamma_{is}(0) = 0$ and $\phi_{is} = 0$ with other settings unchanged. 
As shown in Fig.~\ref{LFCTrackingError-Coma} and Fig.~\ref{LFCEstimationError-Coma}, under this modification, neither the tracking error nor the state estimation errors converge to zero.

\section{Distributed Observer-Based Sliding Mode Control}\label{SDOBSMC}
\subsection{Theoretical Results}
For the case where system \eqref{LTIsys} is subject to matched unknown inputs, consider sliding mode state feedback control law \eqref{centralsliding}, where the gain $\beta$ satisfies $\beta  \ge \bar v\left\| {{X_v}} \right\|$, and $K$ and $P$ satisfy the following inequality:
\begin{equation*}
	(A+BK)^\top P + P (A+BK) < 0.
\end{equation*}
It can be verified, by using Lyapunov function $V = x^\top P x$, that applying \eqref{centralsliding} can stabilize system \eqref{LTIsys}.

Due to \eqref{Busplit}, there exists a matrix $\tilde B$ such that
$$B = \left[ {\begin{array}{*{20}{c}}
		{{B^1}}&{{B^2}}& \cdots &{{B^{{N_c}}}}
\end{array}} \right]\tilde B.$$
Consequently, there exist matrices $K^\iota,\ \iota=1,2,\cdots,N_c$, such that
\begin{equation}\label{lyainequsliding}
	{\left(A + \sum\limits_{\iota  = 1}^{N_c} {{B^\iota }{K^\iota }}\right)^{\top}}P + P\left(A + \sum\limits_{\iota  = 1}^{N_c} {{B^\iota }{K^\iota }}\right) < 0.
\end{equation}

\begin{thm} \label{smthm}
	Under Assumptions~\ref{vbound}, \ref{controllable}, \ref{matchcondition}, \ref{ooconnected}, and \ref{collecdetec}, distributed observer-based sliding mode control law 
	\begin{equation}\label{idealslidinglaw} 
		u^\iota = K^\iota \hat x_{i^\prime} - \beta^\iota h\left({(B^\iota)^{\top}}P \hat x_{i^\prime}\right),\ \iota=1,2,\cdots,N_c
	\end{equation}
	with scalar gains satisfying $\beta^\iota \ge \bar v\left\|\tilde B {{X_v}} \right\|$, $\hat x_{i^\prime}$ the state estimate provided by any observer node, and $K^\iota$ and $P$ satisfying \eqref{lyainequsliding}, can stabilize system \eqref{LTIsys}, i.e., achieve
	\begin{equation*}
		\mathop {\lim }\limits_{t \to \infty } \left\| {x(t)} \right\| = 0.
	\end{equation*}
	Moreover, the state estimation error $e_i$ converges to zero, and adaptive gains $\gamma_i$ and $\gamma_{is}$ remain bounded, $\forall i \in \mathcal{N}$.
\end{thm}

The proof of Theorem~\ref{smthm} is omitted from the paper, since it is similar to that of Theorem~\ref{adsmthm}.

Prior knowledge of the bound of the unknown input is needed for calculating the threshold of $\beta^\iota$ in \eqref{idealslidinglaw}.
Moreover, implementing \eqref{idealslidinglaw} introduces high-frequency switching due to the discontinuous function \eqref{hfunc}.
To alleviate this phenomenon, a boundary layer technique \cite{C.Edwards1998,G.Wheeler1997} can be employed to smooth the control inputs, which leads to the following theorem. 

\begin{thm} \label{adsmthm}
	Under Assumptions~\ref{vbound}, \ref{controllable}, \ref{matchcondition}, \ref{ooconnected}, and \ref{collecdetec}, consider carrying out distributed observer-based sliding mode control law 
	\begin{equation*} \label{adaptivesliding}
		u^\iota = K^\iota \hat x_{i^\prime} - \beta^\iota h^\iota_{\epsilon}\left({(B^\iota)^{\top}}P \hat x_{i^\prime}\right),\ \iota=1,2,\cdots,N_c,
	\end{equation*}
	where $K^\iota$ and $P$ satisfy \eqref{lyainequsliding}; ${h^\iota_{\epsilon}}(\cdot)$ is designed as
	\begin{equation} \label{hepsfunc}
		{h^\iota_{\epsilon}}(\omega) = 
		\left\{
		\begin{aligned}
			{\left\| \omega \right\|}^{-1}{\omega},\ \beta^\iota \left\|\omega\right\| &> \epsilon \\
			\beta^\iota \epsilon^{-1}\omega,\ \beta^\iota \left\|\omega\right\| &\le \epsilon; 
		\end{aligned}
		\right.
	\end{equation}
	$\beta^\iota$, as well as $\gamma_i$ and $\gamma_{is}$ in \eqref{H_idefine}, is updated according to
	\begin{subequations}\label{gagasbeta}
	\begin{align}
		{{\dot \beta }^\iota} &=  - {\sigma ^\iota}{\beta ^\iota} + {\phi ^\iota}\left\| {{{({B^\iota})}^ \top }P{{\hat x}_{i^\prime}}} \right\| \label{adlawbeta}\\
		{{\dot \gamma }_i} &=  - {\sigma _i}{\gamma _i} + {\phi _i}{\left\| {{\varepsilon _{iu}}} \right\|^2} \\
		{{\dot \gamma }_{is}} &=  - {\sigma _{is}}{\gamma _{is}} + {\phi _{is}}\left\| {{\varepsilon _{iu}}} \right\|
	\end{align}
	\end{subequations}
	with leakage coefficients $\sigma^\iota$, $\sigma_i$, $\sigma_{is}$, step sizes ${\phi^{\iota}}$, ${\phi_{i}}$, ${\phi_{is}}$, and initial values $\beta^\iota(0)$, $\gamma_i(0)$, $\gamma_{is}(0)$ all positive reals; $\hat x_{i^\prime}$ is produced by any observer node designed in Section~\ref{SFDOWUGI}.
Then adaptive gains $\beta^\iota$, $\gamma_i$, and $\gamma_{is}$ remain bounded. Moreover, state $x$ of system \eqref{LTIsys} and state estimation error $e_i$ at each observer node exponentially converge to a set containing the origin. 
This residual set can be made arbitrarily small by decreasing 
$\epsilon$ in \eqref{hepsfunc} and by increasing $\phi^\iota$, $\phi_{i}$, and $\phi_{is}$ in \eqref{gagasbeta}.
\end{thm}
See Section~\ref{proofadsmthm} for the proof of Theorem~\ref{adsmthm}.

\begin{rem}
	If $N_c = N$, $\tilde{B} =I$, and the $i$th controller node uses the state estimate from the $i$th observer node, then the set in Theorem~\ref{adsmthm} is given by \eqref{residualset} and the convergence rate is no slower than $\min \{ \sigma ,{\sigma _d}\}$.
\end{rem}

\begin{rem}
	The introduction of the leakage terms with positive coefficients in \eqref{gagasbeta} arises from two fundamental considerations:
\begin{itemize}
	\item As pointed out in \cite{G.Wheeler1997} and \cite{Z.Li2014}, applying boundary layer technique may lead to parameter-drift problems. Without the leakage terms, the adaptive gain $\beta^\iota$ will keep increasing and diverge to infinity.
	\item The adaptive law \eqref{adlawbeta} relies on the estimated state rather than the true state of system \eqref{LTIsys}. 
	Due to the unavoidable estimation errors in the initial stage, the absence of leakage terms may compromise stability of the overall adaptive system.
\end{itemize}
\end{rem}

\subsection{Numerical Example}\label{NE3}
\begin{figure}[htpb]\centering
	\centerline{\includegraphics[width=0.45\textwidth]{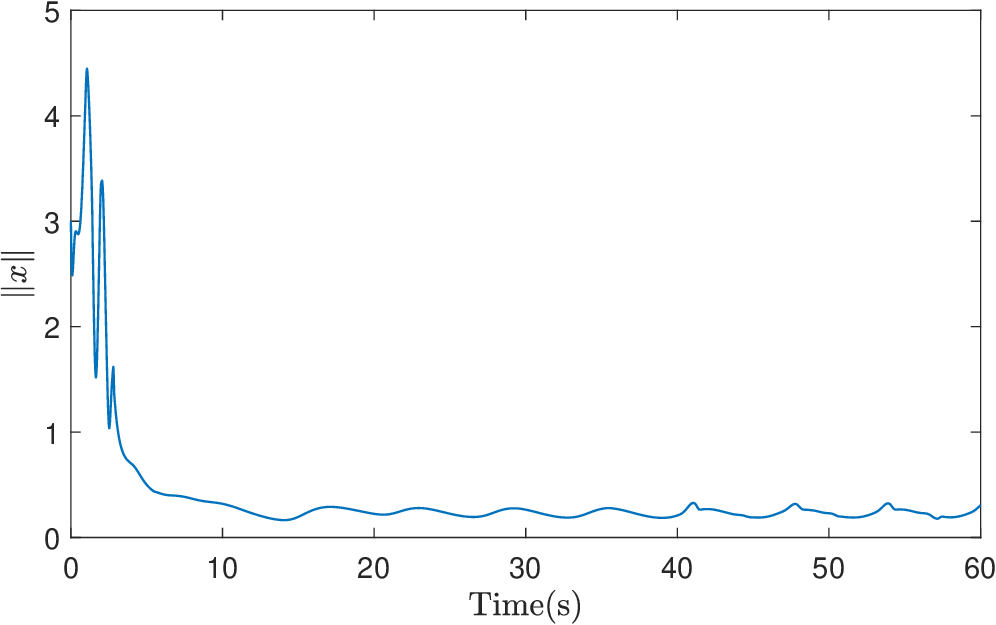}}
	\caption{Stabilization error norm in Section~\ref{NE3}.}
	\label{MSMCStablizationError}
\end{figure}

\begin{figure}[htpb]\centering
	\centerline{\includegraphics[width=0.45\textwidth]{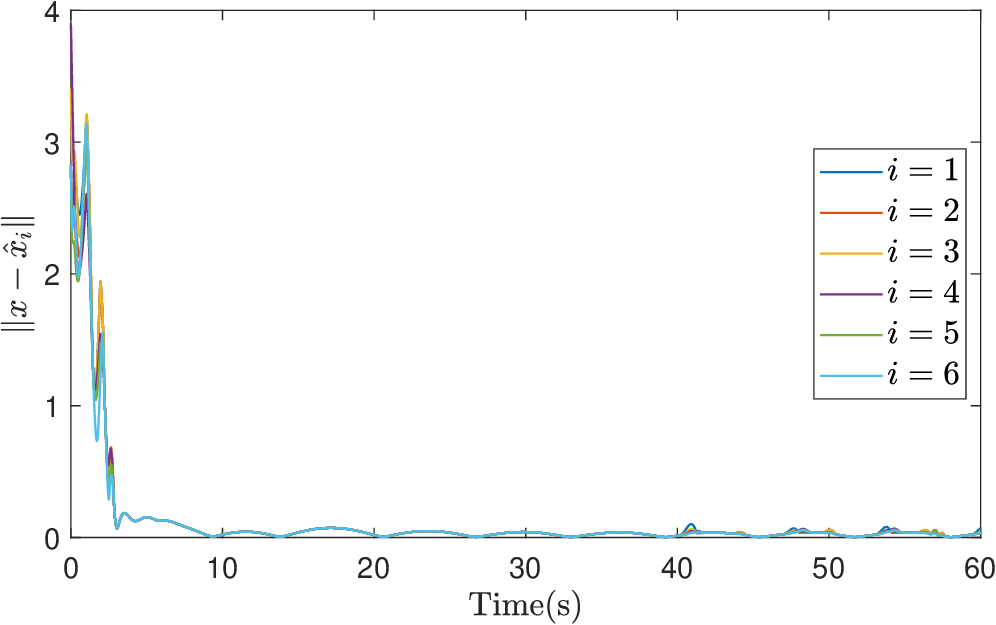}}
	\caption{State estimation error norms in Section~\ref{NE3}.}
	\label{MSMCEstimationError}
\end{figure}

\begin{figure}[htpb]\centering
	\centerline{\includegraphics[width=0.45\textwidth]{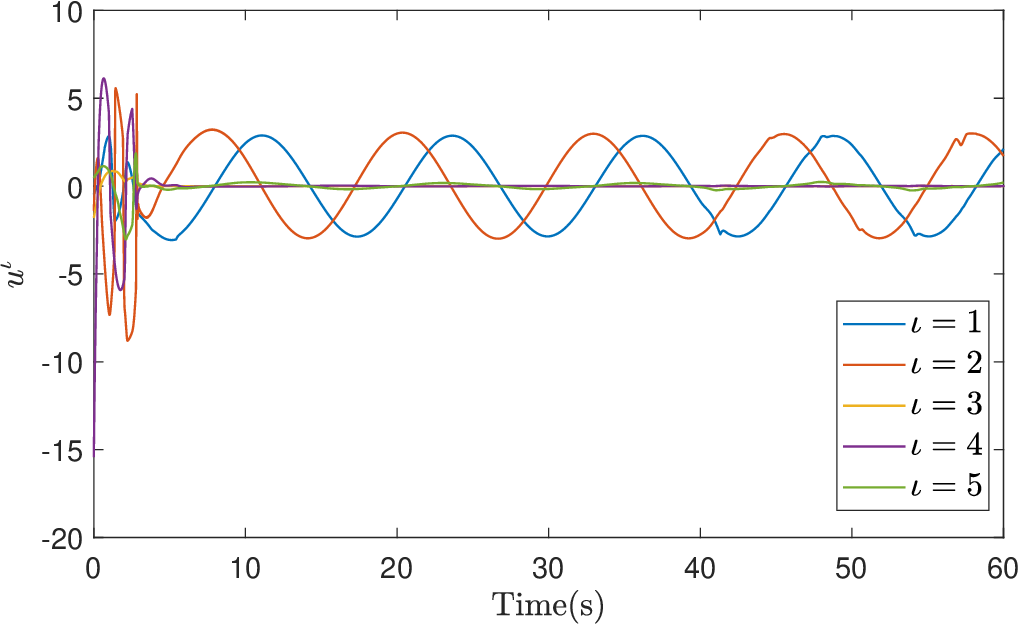}}
	\caption{Control inputs in Section~\ref{NE3}.}
	\label{MSMCControlInput}
\end{figure}

\begin{figure}[htpb]\centering
	\centerline{\includegraphics[width=0.45\textwidth]{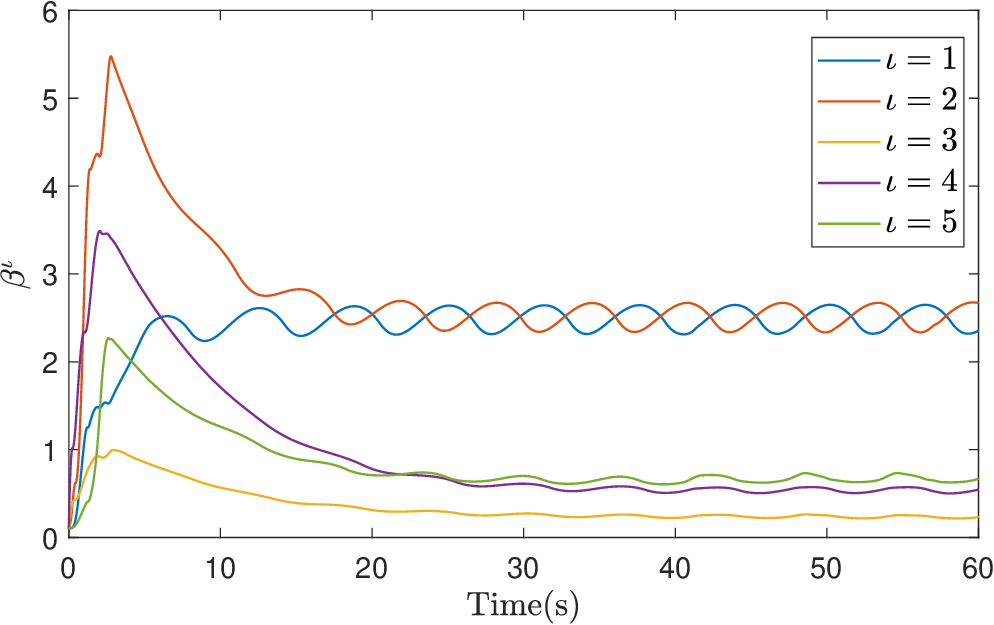}}
	\caption{Adaptive gains in the distributed controllers in Section~\ref{NE3}.}
	\label{MSMCbeta}
\end{figure}

Consider the stabilization problem for system \eqref{LTIsys}, where the unknown input $v(t)$ is generated by the following system:
\begin{equation*}
	\dot v(t) = \begin{bmatrix} 
		0&0.5\\
		-0.5&0
	\end{bmatrix} v(t),\ v(0) = \begin{bmatrix} 
	-2\\
	2
	\end{bmatrix}.
\end{equation*}
Let $A$, $B$, $B^\iota$, $K^\iota$, the number of controller/observer nodes, and C-O/O-O links be the same as those in Section~\ref{NE2}. In addition, set
\begin{align*}
	\beta^\iota(0)=&0.1,\ \sigma^\iota = 0.1,\ \phi^\iota = 5,\ \iota=1,2,3,4,5 \\
	\gamma_i(0)=&0.1,\ \sigma_i = 0.2,\ \phi_i = 5 \\
	\gamma_{is}(0)=&0.1,\ \sigma_{is} = 0.1,\ \phi_{is} = 10,\ i=1,2,3,4,5,6.
\end{align*}
Choose $P$ as the unique positive definite solution of the following Lyapunov equation:
\begin{equation*}
	(A + BK)^\top P + P(A + BK) = -I,
\end{equation*}
where $K = {\rm{col}}({K^\iota})_{\iota = 1}^5$. Set $\epsilon = 0.2$, $B_v = \begin{bmatrix} 
	B^1&B^2
\end{bmatrix}$ and 
\begin{align*}
	x(0) =&
	\begin{bmatrix} 
		1&-1&1&-1&1&-1&1&-1&1
	\end{bmatrix}^\top.
\end{align*}
By using the algorithm provided in Section~\ref{appendix-FUIO}, it can be obtained that
\begin{align*}
	T_{2d} =& 
	\begin{bmatrix} 
		0&0&1&0&0&0&0&0&0
	\end{bmatrix}^\top,
\end{align*}
with $T_{1d}$, $T_{3d}$, $T_{4d}$, $T_{5d}$, and $T_{6d}$ identical to those in Section~\ref{NE2}.
For brevity, the values of matrices $E_i$, $F_i$, $G_i$ are not listed. 
The simulation results in Fig.~\ref{MSMCStablizationError} and Fig.~\ref{MSMCEstimationError} demonstrate that both the system state and the state estimation errors converge to a small neighborhood of the origin. As shown in Fig.~\ref{MSMCControlInput}, the control inputs exhibit virtually no chattering. As shown in Fig.~\ref{MSMCbeta}, the adaptive gains in the distributed controller remain bounded throughout the simulation.


\section{Conclusion}
For linear time-invariant multi-channel systems, this paper presented a fully distributed observer architecture that operates without requiring global control inputs.
Based on this foundation, several forms of distributed controllers were introduced. In the absence of unknown inputs, any linear state-feedback stabilizing controller can be implemented directly by replacing the true state with the estimated state from any observer node. This result was extended to address tracking problems, that is, to track the state trajectory of a reference system.
In the presence of matched unknown inputs, a distributed sliding mode control design was formulated to accomplish stabilization, through which the system states can be driven to zero. To mitigate chattering phenomena of the controller, a boundary layer technique was further incorporated, and meanwhile, an adaptive algorithm was integrated to eliminate the need for prior bounds on unknown inputs.
Theoretical and simulation results demonstrated that the system state can be steered into a small neighborhood around the origin by using virtually smooth control inputs.

A promising future research direction is to bring out advanced collaborative or competitive behaviors in heterogeneous multi-agent systems, based on the fully distributed observers developed in the current work.

\section{Appendix}\label{appendix}
\subsection{Algorithm Associated with Lemma~\ref{FUIO}} \label{appendix-FUIO}
	\textit{Step 1:} Find nonsingular matrices ${\varPhi _0}$ and ${\varPsi _0}$ such that ${\varPhi _0}\left[ {\begin{array}{*{20}{c}}
			0_{p_i \times m_{-i}}\\
			{C_i B_{-i}}
	\end{array}} \right]{\varPsi _0} = \left[ {\begin{array}{*{20}{c}}
			{{I_r}}&0\\
			0&0
	\end{array}} \right]$, where $r$ is the rank of $C_i B_{-i}$. 
	
	\textit{Step 2:} Partition the matrices $B_{-i} {\varPsi_0}$ and ${\varPhi_0}\left[ {\begin{array}{*{20}{c}}
			{{C_i}}\\
			{{C_i}A}
	\end{array}} \right]$ as $B_{-i}{\varPsi_0} = \left[ {\begin{array}{*{20}{c}}
			{{\varLambda _{1,1}}}&{{\varLambda _{1,2}}}
	\end{array}} \right]$ and ${\varPhi _0}\left[ {\begin{array}{*{20}{c}}
	{{C_i}}\\
	{{C_i}A}
\end{array}} \right] = \left[ {\begin{array}{*{20}{c}}
			{{\varPi _1}}\\
			{{\varPi _2}}
	\end{array}} \right]$, where ${\varLambda _{1,1}}$ has $r$ columns and ${\varPi _1}$ has $r$ rows. Find a nonsingular matrix ${\varPsi _1}$ such that ${\varXi _1}{\varPsi _1} = \left[ {\begin{array}{*{20}{c}}
			{{I_{{r_1}}}}&0
	\end{array}} \right]$, where ${\varXi _1}$ is the annihilator of ${\varLambda _{1,2}}$.
	
	\textit{Step 3:} Partition the matrix $\left[ {\begin{array}{*{20}{c}}
			{{\varXi _1}(A - {\varLambda _{1,1}}{\varPi _1})}\\
			{ - {\varPi _2}}
	\end{array}} \right]{\varPsi _1}$ as 
	$\left[ {\begin{array}{*{20}{c}}
			{{\varXi _1}(A - {\varLambda _{1,1}}{\varPi _1})}\\
			{ - {\varPi _2}}
	\end{array}} \right]{\varPsi _1} = \left[ {\begin{array}{*{20}{c}}
			{{\varLambda _{2,1}}}&{{\varLambda _{2,2}}}
	\end{array}} \right]$, where ${\varLambda _{2,1}}$ has ${r_1}$ columns. Partition ${\varXi _2}$ as ${\varXi _2} = \left[ {\begin{array}{*{20}{c}}
			{{\varXi _{2,1}}}&{{\varXi _{2,2}}}
	\end{array}} \right]$, where ${\varXi _2}$ is the annihilator of ${\varLambda _{2,2}}$ and ${\varXi _{2,1}}$ has $r_1$ columns. Initialize the iteration index $j=2$ for the subsequent iterations.
	
	\textit{Step 4:} If ${\varXi _{j,1}}$ is of full column rank, proceed directly to \textit{Step 6}. If ${\varXi _{j,1}}$ is a zero matrix or does not exist (i.e., ${\varLambda _{j,2}}$ has full row rank), let $r_d=0$ and skip to \textit{Step 7}. Otherwise, proceed to \textit{Step 5}.
	
	\textit{Step 5:} Find nonsingular matrices ${\varPhi _j}$ and ${\varPsi _j}$ such that ${\varPhi _j}{\varXi _{j,1}}{\varPsi _j} = \left[ {\begin{array}{*{20}{c}}
			{{I_{{r_j}}}}&0\\
			0&0
	\end{array}} \right]$. Partition ${\varPhi _j}{\varXi _j}{\varLambda _{j,1}}{\varPsi _j}$ as 
	${\varPhi _j}{\varXi _j}{\varLambda _{j,1}}{\varPsi _j} = \left[ {\begin{array}{*{20}{c}}
			{{\varLambda _{j + 1,1}}}&{{\varLambda _{j + 1,2}}}
	\end{array}} \right]$, where ${\varLambda _{j + 1,1}}$ has ${r_j}$ columns. Partition ${\varXi _{j + 1}}$ as ${\varXi _{j + 1}} = \left[ {\begin{array}{*{20}{c}}
			{{\varXi _{j + 1,1}}}&{{\varXi _{j + 1,2}}}
	\end{array}} \right]$, where ${\varXi _{j + 1}}$ is the annihilator of ${\varLambda _{j + 1,2}}$ and ${\varXi _{j + 1,1}}$ has ${r_j}$ columns. Let $j = j + 1$, and return to \textit{Step 4}.
	
	\textit{Step 6:} Partition ${\varPhi _j}{\varXi _j}{\varLambda _{j,1}}$ as ${\varPhi _j}{\varXi _j}{\varLambda _{j,1}} = \left[ {\begin{array}{*{20}{c}}
			{{\varPi _3}}\\
			{{\varPi _4}}
	\end{array}} \right]$, where ${\varPi _3}$ has ${r_j}$ rows. Compute the detectability decomposition \cite{W.M.Wonham1985} of the pair $\left( {{\varPi _3},{\varPi _4}} \right)$ to give 
	\begin{equation*}
		P_\varPi ^{\top}{\varPi _3}{P_\varPi } = \left[ {\begin{array}{*{20}{c}}
				{{\varPi _{3d}}}&0\\
				{{\varPi _{3r}}}&{{\varPi _{3u}}}
		\end{array}} \right]{\rm{, \ }}{\varPi _4}{P_\varPi } = \left[ {\begin{array}{*{20}{c}}
				{{\varPi _{4d}}}&0
		\end{array}} \right],
	\end{equation*} 
	where ${\varPi _{3d}} \in {\mathbb{R}^{{r_d} \times {r_d}}}$. Let $\delta_i  = \max \left\{ {{r_d},{p_i}} \right\}$, where ${p_i}$ is the rank of $C_i$.
	
	\textit{Step 7:} If ${r_d} > {p_i}$, let $\varPhi  = {\varPhi _j}{\varXi _j}{\varPhi _{j - 1}}{\varXi _{j - 1}} \cdots {\varPhi _3}{\varXi _3}{\varPhi _2}$ and ${J_1} = \left[ {\begin{array}{*{20}{c}}
			{{I_{{r_d}}}}&{{0_{{r_d} \times ({r_j} - {r_d})}}}
	\end{array}} \right]P_\varPi ^{\top}$, and find a matrix ${J_2}$ such that ${\varPi _{3d}} + {J_2}{\varPi _{4d}}$ is stable. Then choose 
	\begin{equation*}
		\begin{split}
			{E_{i0}} &= {\varPi _{3d}} + {J_2}{\varPi _{4d}}\\
			T_{id0}^{\top} &= J\varPhi {\varXi _{2,1}}{\varXi _1} \\
			{K_{i0}} &= \left[ {\begin{array}{*{20}{c}}
					{T_{id0}^{\top}{\varLambda _{1,1}}}&{J\varPhi {\varXi _{2,2}}}
			\end{array}} \right]{\varPhi _0},
		\end{split}
	\end{equation*}
	where $J = \left[ {\begin{array}{*{20}{c}}
			{{J_1}}&{{J_2}}
	\end{array}} \right]$.
	If ${r_d} \le {p_i}$, choose ${E_{i0}} = {0_{{p_i} \times {p_i}}}$, $T_{id0}^{\top} = {C_i}$, and ${K_{i0}} = \left[ {\begin{array}{*{20}{c}}
			{{0_{{p_i} \times {p_i}}}}&I_{p_i}
	\end{array}} \right]$.

	\textit{Step 8:} Perform a Gram–Schmidt orthonormalization ${\tilde T_{id}}T_{id0}^{\top} = T_{id}^{\top}$, where ${\tilde T_{id}}$ is a nonsingular matrix, and the columns of ${T_{id}}$ are orthonormal. Then the matrix quadruplet $\left( {{T_{id}},{E_i},{F_i},{G_i}} \right)$ in Lemma~\ref{FUIO} is given by
	\begin{equation*}
		\begin{split}
			{T_{id}} &= {T_{id0}}\tilde T_{id}^{\top},\ {E_i} = {{\tilde T}_{id}}{E_{i0}}\tilde T_{id}^{ - 1}\\
			G_i&= {{\tilde T}_{id}}{K_{i0}}\left[ {\begin{array}{*{20}{c}}
					0_{{p_i} \times {p_i}}\\
					I_{p_i}
			\end{array}} \right]\\
		    F_i&=E_i G_i + {{\tilde T}_{id}}{K_{i0}}\left[ {\begin{array}{*{20}{c}}
		    		I_{p_i} \\
		    		0_{{p_i} \times {p_i}}
		    \end{array}} \right].
		\end{split}
	\end{equation*}

\subsection{Proof of Theorem~\ref{observerthm}} \label{proofobserverthm}
Define the state estimation error at the $i$th node as 
\begin{equation}\label{eidefine}
	{e_i} = {{\hat x}_i} - x,
\end{equation}
and examine the dynamics of $T_{id}^{\top}{e_i}$.
By taking advantage of \eqref{localmeasure}, \eqref{obidynamic}, \eqref{barEFGB}, \eqref{TidTiu}, \eqref{H_idefine}, \eqref{eidefine}, it is obtained that 
\begin{align}
	T_{id}^{\top}{{\dot e}_i} =& T_{id}^{\top}({{\dot z}_i} + {{\bar G}_i}{C_i}\dot x) - T_{id}^{\top}\dot x \nonumber\\
	=& {E_i}T_{id}^{\top}{z_i} + {F_i}{y_i} + T_{id}^{\top}{{\bar B}_i}{u_i} 
	+ ({G_i}{C_i} - T_{id}^\top){\dot x}.\label{Tiddotei1}
\end{align}
It follows from \eqref{hatxieq}, \eqref{barG_i}, \eqref{TidTiu}, \eqref{eidefine} that
\begin{equation} \label{Tidzi}
	T_{id}^{\top}{z_i} = T_{id}^{\top}{e_i} - {G_i}{y_i} + T_{id}^{\top}x.
\end{equation}
Substituting \eqref{LTIsys}, \eqref{splitinput}, and \eqref{Tidzi} into \eqref{Tiddotei1} yields
\begin{align*}
	T_{id}^{\top}{{\dot e}_i} =& {E_i}T_{id}^{\top}{e_i} + ({F_i} - {E_i}{G_i}){y_i} + ({E_i}T_{id}^{\top} + {G_i}{C_i}A - T_{id}^{\top}A)x \nonumber\\
	& + (T_{id}^{\top}{{\bar B}_i} + {G_i}{C_i}{B_i} - T_{id}^{\top}{B_i}){u_i} + ({G_i}{C_i} - T_{id}^{\top}){B_{ - i}}{u_{ - i}}.
\end{align*}
Based on \eqref{localmeasure}, \eqref{barB_i}, \eqref{TidTiu}, and Lemma~\ref{FUIO}, we obtain that 
\begin{equation*}
	T_{id}^{\top}{{\dot e}_i} = {E_i}T_{id}^{\top}{e_i}.
\end{equation*}
In particular, when $E_i = 0$, the algorithm associated with Lemma~\ref{FUIO} gives $G_i C_i = T_{id}^\top$ and $F_i = 0$. Then it follows from \eqref{zidynamic}, \eqref{barEFGB}, \eqref{H_idefine} that $T_{id}^\top \dot z_i = 0$.
Finally, according to \eqref{hatxieq}, \eqref{barG_i}, \eqref{eidefine}, it follows that $T_{id}^{\top}{e_i} \equiv 0.$

Next, look into the dynamics of $T_{iu}^{\top}{e_i}$. From \eqref{LTIsys}, \eqref{splitinput}, \eqref{obidynamic}, \eqref{barEFGB}, \eqref{TidTiu}, \eqref{H_idefine}, \eqref{eidefine}, it follows that 
\begin{align}
	T_{iu}^{\top}{{\dot e}_i} =& T_{iu}^{\top}A{z_i} + T_{iu}^{\top}A{{\bar G}_i}{y_i} + T_{iu}^{\top}{{\bar B}_i}{u_i} - {\gamma _i}{\varepsilon _{iu}} - {\gamma _{is}}h({\varepsilon _{iu}}) \nonumber\\
	&- T_{iu}^{\top}(Ax + {B_i}{u_i} + {B_{ - i}}{u_{ - i}}).\nonumber
\end{align}
Then according to \eqref{hatxieq}, \eqref{eidefine}, \eqref{barB_i}, it is further obtained
\begin{align}
	T_{iu}^{\top}{{\dot e}_i} =& T_{iu}^{\top}A{{\hat x}_i} - T_{iu}^{\top}Ax - T_{iu}^{\top}{B_{ - i}}{u_{ - i}} \nonumber\\
	& + (T_{iu}^{\top}{{\bar B}_i} - T_{iu}^{\top}{B_i}){u_i} - {\gamma _i}{\varepsilon _{iu}} - {\gamma _{is}}h({\varepsilon _{iu}})\nonumber\\
	=& T_{iu}^{\top}A{e_i} - T_{iu}^{\top}{B_{ - i}}{u_{ - i}} - {\gamma _i}{\varepsilon _{iu}} - {\gamma _{is}}h({\varepsilon _{iu}}). \nonumber
\end{align}
With relation \eqref{TidTiu}, we can finally arrive at
\begin{align} \label{Tiueidyna}
	T_{iu}^{\top}{{\dot e}_i} =& T_{iu}^{\top}A{T_{iu}}T_{iu}^{\top}{e_i} + T_{iu}^{\top}A{T_{id}}T_{id}^{\top}{e_i} - T_{iu}^{\top}{B_{ - i}}{u_{ - i}} \nonumber\\
	&- {\gamma _i}{\varepsilon _{iu}} - {\gamma _{is}}h({\varepsilon _{iu}}).
\end{align}
Let ${\varepsilon _{id}} = T_{id}^{\top}{e_i}$, and rewrite $\varepsilon_{iu} =T_{iu}^\top \sum\nolimits_{j = 1}^N {a_{ij} (\hat x_i - \hat x_j)}$ as
\begin{align}
	{\varepsilon _{iu}} =& T_{iu}^{\top}\sum\limits_{j = 1}^N {{a_{ij}}({e_i} - {e_j})} \nonumber\\
	=&T_{iu}^{\top}\sum\limits_{j = 1}^N {{l_{ij}}({T_{id}}T_{id}^{\top} + {T_{iu}}T_{iu}^{\top}){e_j}}.\nonumber
\end{align}
Then it can be verified that the following relation holds
\begin{equation}\label{nonsintrans}
	\left[ {\begin{array}{*{20}{c}}
			{{\varepsilon _d}}\\
			{{\varepsilon _u}}
	\end{array}} \right] = \left[ {\begin{array}{*{20}{c}}
			I_{\sum\nolimits_{i = 1}^N {{\delta _i}} }&0\\
			{T_u^{\top}({\cal L} \otimes I_n){T_d}}&{T_u^{\top}({\cal L} \otimes I_n){T_u}}
	\end{array}} \right]\left[ {\begin{array}{*{20}{c}}
			{T_d^{\top}e}\\
			{T_u^{\top}e}
	\end{array}} \right],
\end{equation}
where ${\varepsilon _d} = {\rm{col}}({\varepsilon _{id}})_{i = 1}^N$, ${\varepsilon _u} = {\rm{col}}({\varepsilon _{iu}})_{i = 1}^N$, $e = {\rm{col}}({e_i})_{i = 1}^N$, ${T_d} = {\rm{diag}}({T_{id}})_{i = 1}^N$, and ${T_u} = {\rm{diag}}({T_{iu}})_{i = 1}^N$.
Moreover, the matrix ${T_u^{\top}({\cal L} \otimes I_n){T_u}}$ in \eqref{nonsintrans} is positive definite according to \eqref{TidTiu}, Assumptions~\ref{ooconnected} and \ref{collecdetec}, and Lemma~\ref{PosiDef}.
Given that $\left[ {\begin{array}{*{20}{c}}
		{{T_d}}&{{T_u}}
\end{array}} \right]$ is also a nonsingular matrix, it suffices to check the stability of ${\varepsilon_d}$ and ${\varepsilon_u}$, instead of considering that of $e$. 

The dynamics of ${\varepsilon_d}$ are of the form
\begin{equation} \label{dotepsd}
	\dot\varepsilon_d = E\varepsilon_d,
\end{equation}
where $E={\rm{diag}}({E_i})_{i = 1}^N$ with $E_i$ being either a stable or zero matrix. Given that $\varepsilon_{id}\equiv 0$ if $E_i = 0$, it follows that $\varepsilon_d$ exponentially converges to zero with time.

With the aid of \eqref{Tiueidyna}, \eqref{nonsintrans}, \eqref{dotepsd}, and the relation
\begin{equation}\label{Tueepsud}
	T_u^{\top}e = {\left[ {T_u^{\top}({\cal L} \otimes {I_n}){T_u}} \right]^{ - 1}}\left[ {{\varepsilon _u} - T_u^{\top}({\cal L} \otimes {I_n}){T_d}{\varepsilon _d}} \right],
\end{equation}
the dynamics of ${\varepsilon_u}$ can be expressed as 
\begin{align} 
	{{\dot \varepsilon }_u} =& T_u^{\top}({\cal L} \otimes I){T_d}{{\dot \varepsilon }_d} + T_u^{\top}({\cal L} \otimes I){T_u}T_u^{\top}\dot e \nonumber\\
	=& {{\bar A}_r}{\varepsilon _d} + {{\bar A}_u}{\varepsilon _u} - T_u^{\top}({\cal L} \otimes I){T_u}(\Gamma {\varepsilon _u} + T_u^{\top}{B_ - }{u_ - } + {\Gamma _s}{\bar h}), \label{dotepsu}
\end{align}
where $\Gamma  = {\rm{diag}}({\gamma _i}{I_{n - {\delta _i}}})_{i = 1}^N$, ${\Gamma _s} = {\rm{diag}}({\gamma _{is}}{I_{n - {\delta _i}}})_{i = 1}^N$, ${B_ - }{u_ - } = {\rm{col}}({B_{ - i}}{u_{ - i}})_{i = 1}^N$,
\begin{align*}
	{\bar h} =& {\left[ {\begin{array}{*{20}{c}}
				{{h^{\top}}({\varepsilon _{1u}})}&{{h^{\top}}({\varepsilon _{2u}})}& \cdots &{{h^{\top}}({\varepsilon _{Nu}})}
		\end{array}} \right]^{\top}} \nonumber\\
	{{\bar A}_u} =& T_u^{\top}({\cal L} \otimes I){T_u}T_u^{\top}(I \otimes A){T_u}{\left[ {T_u^{\top}({\cal L} \otimes I){T_u}} \right]^{ - 1}} \nonumber\\
	{{\bar A}_r} =& T_u^{\top}({\cal L} \otimes I){T_d}E - {{\bar A}_u}T_u^{\top}({\cal L} \otimes I){T_d} \nonumber\\
	&+ T_u^{\top}({\cal L} \otimes I){T_u}T_u^{\top}(I \otimes A){T_d}. \nonumber
\end{align*}

Choose the Lyapunov function 
\begin{align}
	{V_\varepsilon} =& \frac{1}{2}\varepsilon _u^{\top}{\left[ {T_u^{\top}({\cal L} \otimes I){T_u}} \right]^{ - 1}}{\varepsilon _u} + \sum\limits_{i = 1}^N {\frac{1}{{2{\phi _i}}}{{({\gamma _i} - {\gamma ^*})}^2}} \nonumber \\
	&+ \sum\limits_{i = 1}^N {\frac{1}{{2{\phi _{is}}}}{{({\gamma _{is}} - \gamma_s^*)}^2}}, \label{V_o} 
\end{align}
where ${\gamma^*}$ and $\gamma_s^*$ are two positive constants to be determined later.
Differentiating \eqref{V_o} along the trajectory of \eqref{dotepsu} gives
\begin{align}
	{{\dot V}_\varepsilon } =& \varepsilon _u^{\top}{\left[ {T_u^{\top}({\cal L} \otimes I){T_u}} \right]^{ - 1}}{{\dot \varepsilon }_u} + \sum\limits_{i = 1}^N {\frac{1}{{{\phi _i}}}({\gamma _i} - {\gamma ^*}){{\dot \gamma }_i}}  \nonumber\\
	&+ \sum\limits_{i = 1}^N {\frac{1}{{{\phi _{is}}}}({\gamma _{is}} - \gamma _s^*){{\dot \gamma }_{is}}} \nonumber\\
	=& \varepsilon _u^{\top}{\left[ {T_u^{\top}({\cal L} \otimes I){T_u}} \right]^{ - 1}}({{\bar A}_r}{\varepsilon _d} + {{\bar A}_u}{\varepsilon _u}) - \varepsilon _u^{\top}(\Gamma {\varepsilon _u} + T_u^{\top}{B_ - }{u_ - })  \nonumber\\
	&- \varepsilon _u^{\top}{\Gamma _s}{\bar h} + \sum\limits_{i = 1}^N {({\gamma _i} - {\gamma ^*})\varepsilon _{iu}^{\top}} {\varepsilon _{iu}} + \sum\limits_{i = 1}^N {({\gamma _{is}} - \gamma _s^*)\left\| {{\varepsilon _{iu}}} \right\|} \nonumber\\
	=& \varepsilon _u^{\top}{\left[ {T_u^{\top}({\cal L} \otimes I){T_u}} \right]^{ - 1}}({{\bar A}_r}{\varepsilon _d} + {{\bar A}_u}{\varepsilon _u}) - \varepsilon _u^{\top}T_u^{\top}{B_ - }{u_ - } \nonumber\\
	& - {\gamma ^*}\sum\limits_{i = 1}^N {\varepsilon _{iu}^{\top}{\varepsilon _{iu}}}  - \gamma _s^*\sum\limits_{i = 1}^N {\left\| {{\varepsilon _{iu}}} \right\|}.	\nonumber
\end{align}
Using the inequality
\begin{equation*}
	\varepsilon _u^{\top}{\left[ {T_u^{\top}({\cal L} \!\otimes\! I){T_u}} \right]^{\! - 1}}\!\!{{\bar A}_r}{\varepsilon _d} \!\le\! \frac{1}{4}{\left\| {{{\left[ {T_u^{\top}({\cal L} \!\otimes\! I){T_u}} \right]}^{\! - 1}}\!\!{{\bar A}_r}} \right\|^2}\varepsilon _u^{\top}{\varepsilon _u} + \varepsilon _d^{\top}{\varepsilon _d},
\end{equation*}
it then follows that
\begin{align*}
	{{\dot V}_\varepsilon } \le& ({\lambda _a} + {\lambda _b} - {\gamma ^*})\varepsilon _u^{\top}{\varepsilon _u} + \varepsilon _d^{\top}{\varepsilon _d} + {\lambda _c}\left\| {{\varepsilon _u}} \right\| - \gamma _s^*\sum\limits_{i = 1}^N {\left\| {{\varepsilon _{iu}}} \right\|} \nonumber\\
	\le& \varepsilon _d^{\top}{\varepsilon _d} + ({\lambda _a} + {\lambda _b} - {\gamma ^*})\varepsilon _u^{\top}{\varepsilon _u} + ({\lambda _c} - \gamma _s^*)\left\| {{\varepsilon _u}} \right\|,
\end{align*} 
where 
\begin{subequations}\label{lambdaabc}
	\begin{align}
		{\lambda _a} =& \frac{1}{4}{\left\| {{{\left[ {T_u^{\top}({\cal L} \otimes I){T_u}} \right]}^{ - 1}}{{\bar A}_r}} \right\|^2} \\
		{\lambda _b} =& \left\| {{{\left[ {T_u^{\top}({\cal L} \otimes I){T_u}} \right]}^{ - 1}}{{\bar A}_u}} \right\| \\
		{\lambda _c} =& \left\| {T_u^{\top}{B_ - }{u_ - }} \right\|.
	\end{align}
\end{subequations}
Selecting $\gamma^* > {\lambda _a} + {\lambda _b}$ and $\gamma_s^* > {\lambda _c}$ guarantees that
\begin{equation}\label{dotVeps-epsd}
	\dot V_{\varepsilon} - \varepsilon_d^{\top}{\varepsilon _d} \le({\lambda _a} + {\lambda _b} - {\gamma ^*})\varepsilon _u^{\top}{\varepsilon _u} \le 0,
\end{equation}  
which implies that $V_{\varepsilon}(t) - \int_0^t \varepsilon_d^{\top}(\tau){\varepsilon_d}(\tau) \rm{d}\tau$ is nonincreasing.
Since $\varepsilon_d$ exponentially converges to zero with time, $\int_0^t \varepsilon_d^{\top}(\tau){\varepsilon_d}(\tau) \rm{d}\tau$ exists and remains bounded. Consequently, $V_{\varepsilon}(t)$ is also bounded, which, together with \eqref{V_o}, implies that $\varepsilon_u (t)$, $\gamma_i(t)$, and $\gamma_{is}(t)$ are all bounded.
It then follows from \eqref{dotepsu} and Assumption~\ref{ubound} that $\dot \varepsilon_u (t)$ is bounded, and so $\varepsilon _u^{\top}(t){\varepsilon _u}(t)$ is uniformly continuous. 
Meanwhile, given that $V_{\varepsilon}(t) - \int_0^t \varepsilon_d^{\top}(\tau){\varepsilon_d}(\tau) \rm{d}\tau$ is nonincreasing and has a lower bound, it has a finite limit, i.e., 
$${\lim _{t \to \infty }} \left(V_{\varepsilon}(t) - \int_0^t \varepsilon_d^{\top}(\tau){\varepsilon_d}(\tau) \rm{d}\tau\right) = V_{\varepsilon^\prime}^\infty.$$
Integrating both sides of \eqref{dotVeps-epsd} gives
\begin{equation*}
	\int_0^\infty \left[({{\gamma ^*} -\lambda _a- {\lambda_b} })\varepsilon _u^{\top}(\tau){\varepsilon _u}(\tau)\right] \rm{d}\tau \le V_\varepsilon(0) - V_{\varepsilon^\prime}^\infty,
\end{equation*}
based on which $\int_0^\infty \varepsilon_u^{\top}(\tau){\varepsilon_u}(\tau) \rm{d}\tau$ exists and remains finite.
According to Lemma~\ref{barbalaslemma}, we can arrive at $\lim _{t \to \infty } {\varepsilon_u}(t) = 0$, which completes the proof.

\subsection{Proof of Theorem~\ref{lfthm}} \label{prooflfthm}
It is noted in Theorem~\ref{lfthm} that any controller node may utilize the estimate produced by any observer node.
Moreover, the number of controller nodes and the number of observer nodes are not required to be identical.
To facilitate the proof, however, consider the case where the number of the controller nodes is the same as that of the observer nodes, the $i$th controller node receives and uses the state estimate from the $i$th observer node, and $B = \left[ {\begin{array}{*{20}{c}}
		{{B^1}}&{{B^2}}& \cdots &{{B^N}}
\end{array}} \right]$. 

Under this setting, the control input applied by the $i$th controller node is
\begin{equation}\label{u^isimp}
	u^i = K^i \hat x_i.
\end{equation} 
Furthermore, there exist a positive definite matrix $P$ and a positive constant $\varsigma$ such that
\begin{equation}\label{lyainequ}
	(A+BK)^\top P + P (A+BK) < -\varsigma I,
\end{equation}
where $K = {\rm{col}}({K^i})_{i = 1}^N$.
Choose the Lyapunov function
\begin{align}
	{V_x} =& \frac{1}{2}{x^{\top}}Px + \frac{1}{2}\varepsilon _u^{\top}{\left[ {T_u^{\top}({\cal L} \otimes I){T_u}} \right]^{ - 1}}{\varepsilon _u}  \nonumber \\
	&+ \sum\limits_{i = 1}^N {\frac{1}{{2{\phi _i}}}{{({\gamma _i} - {\gamma ^*})}^2}} + \sum\limits_{i = 1}^N {\frac{1}{{2{\phi _{is}}}}{{({\gamma _{is}} - \gamma_s^*)}^2}}, \label{V_x} 
\end{align}
where ${\gamma^*}$ and $\gamma_s^*$ are two positive constants to be determined later.
After substituting \eqref{Busplit} and \eqref{u^isimp} into \eqref{LTIsys}, differentiating \eqref{V_x} along the trajectories of \eqref{LTIsys} gives
\begin{align}
	{{\dot V}_x} =& {x^{\top}}P(A + BK)x + {x^{\top}}PB{K_D}e +\varepsilon _u^{\top}{\left[ {T_u^{\top}({\cal L} \otimes I){T_u}} \right]^{ - 1}}{{\dot \varepsilon }_u} \nonumber\\
	& + \sum\limits_{i = 1}^N {\frac{1}{{{\phi _i}}}({\gamma _i} - {\gamma ^*}){{\dot \gamma }_i}} + \sum\limits_{i = 1}^N {\frac{1}{{{\phi _{is}}}}({\gamma _{is}} - \gamma _s^*){{\dot \gamma }_{is}}}, \label{dotVx1}
\end{align}
where ${K_D} = {\rm{diag}}({K^i})_{i = 1}^N$. Recall from \eqref{TidTiu}, \eqref{nonsintrans}, \eqref{Tueepsud} that
\begin{align} \label{esplitasepsdu}
	e = {T_d}T_d^{\top}e + {T_u}T_u^{\top}e = {\Delta _d}{\varepsilon _d} + {\Delta _u}{\varepsilon _u},
\end{align}
where 
\begin{align*}
	\Delta_d =& {{T_d} - {T_u}{{\left[ {T_u^{\top}({\cal L} \otimes I){T_u}} \right]}^{ - 1}}T_u^{\top}({\cal L} \otimes I){T_d}} \nonumber\\
	\Delta_u =& {T_u}{\left[ {T_u^{\top}({\cal L} \otimes I){T_u}} \right]^{ - 1}}.
\end{align*}
To proceed with the derivation, express the portion of the control input unavailable to the $i$th observer node as
\begin{equation} \label{u-iexpress}
	{u_{-i}} = {K^{-i}}{{\hat x}_{-i}},
\end{equation}
where ${\hat x}_{-i}$ is a vector formed by the state estimates from corresponding observer nodes, and $K^{-i}$ is a matrix formed by the gains of corresponding controller nodes.
By substituting \eqref{dotepsu}, \eqref{esplitasepsdu}, and \eqref{u-iexpress} into \eqref{dotVx1}, it is obtained that
\begin{align*}
	{{\dot V}_x} \le& {x^{\top}}P(A + BK)x + {x^{\top}}PB{K_D}({\Delta _d}{\varepsilon _d} + {\Delta _u}{\varepsilon _u}) + \varepsilon _d^{\top}{\varepsilon _d} \nonumber\\
	& + ({\lambda _a} + {\lambda _b} - {\gamma ^*})\varepsilon _u^{\top}{\varepsilon _u} - \varepsilon _u^{\top}T_u^{\top}{B_ - }K_D^ - {{\hat x}_- } - \gamma _s^*\left\| {{\varepsilon _u}} \right\|, \nonumber
\end{align*}
where $\lambda_a$ and $\lambda_b$ are defined by \eqref{lambdaabc}, $B_{-} K_D^{-}  = {\rm{diag}}(B_{-i}{K^{ - i}})_{i = 1}^N$, and ${{\hat x}_ - } = {\rm{col}}({{\hat x}_{ - i}})_{i = 1}^N$.
Note that there exists a matrix ${\Delta_-}$ such that
\begin{equation} \label{hatx-xe}
	{{\hat x}_ - } = {\Delta _ - }({1_N} \otimes x + e) = {\Delta _ - }({1_N} \otimes x + {\Delta _d}{\varepsilon _d} + {\Delta _u}{\varepsilon _u}).
\end{equation}
Then from \eqref{lyainequ}, \eqref{hatx-xe}, and the following inequalities
\begin{align*}
	{x^{\top}}PB{K_D}{\Delta _d}{\varepsilon _d} \le& \frac{\varsigma }{8}{x^{\top}}x + \frac{2}{\varsigma }{\left\| {PB{K_D}{\Delta _d}} \right\|^2}\varepsilon _d^{\top}{\varepsilon _d}\nonumber\\
	{x^{\top}}PB{K_D}{\Delta _u}{\varepsilon _u} \le& \frac{\varsigma }{8}{x^{\top}}x + \frac{2}{\varsigma }{\left\| {PB{K_D}{\Delta _u}} \right\|^2}\varepsilon _u^{\top}{\varepsilon _u}\nonumber\\
	- \varepsilon _u^{\top}T_u^{\top}{B_ - }K_D^ - {\Delta _ - }({1_N} \otimes x) \le& \frac{\varsigma }{8}{x^{\top}}x + \frac{{2N}}{\varsigma }{\left\| {T_u^{\top}{B_ - }K_D^ - {\Delta _ - }} \right\|^2}\varepsilon _u^{\top}{\varepsilon _u} \nonumber \\
	- \varepsilon _u^{\top}T_u^{\top}{B_ - }K_D^ - {\Delta _ - }{\Delta _d}{\varepsilon _d} \le& \varepsilon _d^{\top}{\varepsilon _d} + \frac{1}{4}{\left\| {T_u^{\top}{B_ - }K_D^ - {\Delta _ - }{\Delta _d}} \right\|^2}\varepsilon _u^{\top}{\varepsilon _u}, \nonumber
\end{align*}
it follows that 
\begin{align*}
	{{\dot V}_x} \le& -\frac{\varsigma }{8}{x^{\top}}x + \lambda_d \varepsilon _d^{\top}{\varepsilon _d} + ({\lambda_a} + {\lambda_b} + {\lambda_e} - {\gamma ^*})\varepsilon _u^{\top}{\varepsilon _u}- \gamma _s^*\left\| {{\varepsilon _u}} \right\|,
\end{align*}
where
\begin{align*}
	\lambda_d=& 2+ \frac{2}{\varsigma }{\left\| {PB{K_D}{\Delta _d}} \right\|^2}\nonumber\\
	\lambda_e=& \left\| {T_u^{\top}{B_ - }K_D^ - {\Delta _ - }{\Delta _u}} \right\| + \frac{2}{\varsigma }{\left\| {PB{K_D}{\Delta _u}} \right\|^2} \nonumber\\
	&+ \frac{{2N}}{\varsigma }{\left\| {T_u^{\top}{B_ - }K_D^ - {\Delta _ - }} \right\|^2} + \frac{1}{4}{\left\| {T_u^{\top}{B_ - }K_D^ - {\Delta _ - }{\Delta _d}} \right\|^2}.
\end{align*}
Selecting $\gamma^* > {\lambda _a} + {\lambda _b} + {\lambda _e}$ and $\gamma_s^* \ge 0$ guarantees that
\begin{equation}\label{dotVx-lamepsd}
	\dot V_{x} - \lambda_d\varepsilon_d^{\top}{\varepsilon _d} \le -\frac{\varsigma }{8}{x^{\top}}x + ({\lambda _a} + {\lambda _b} + {\lambda _e} - {\gamma ^*})\varepsilon _u^{\top}{\varepsilon _u} \le 0,
\end{equation}
which implies that $V_{x}(t) - \int_0^t \lambda_d \varepsilon_d^{\top}(\tau){\varepsilon_d}(\tau) \rm{d}\tau$ is nonincreasing. Similar to the analysis in Section~\ref{proofobserverthm}, it can be deduced that $x(t)$, $\varepsilon_u (t)$, $\gamma_i(t)$, and $\gamma_{is}(t)$ are all bounded. Furthermore, $\lim _{t \to \infty } {x}(t) = 0$ and $\lim _{t \to \infty } {\varepsilon_u}(t) = 0$ can be concluded.
	
\subsection{Proof of Theorem~\ref{adsmthm}} \label{proofadsmthm}
Without loss of generality, consider the case where the number of the controller nodes is the same as that of the observer nodes, the $i$th controller node takes the state estimate from the $i$th observer node, and $B = \left[ {\begin{array}{*{20}{c}}
		{{B^1}}&{{B^2}}& \cdots &{{B^N}}
\end{array}} \right]$. 
Based on such simplified settings, the $i$th controller node exerts
\begin{equation}\label{u^iadslidsimp}
	u^i = K^i \hat x_{i} - \beta^i h^i_{\epsilon}\left({(B^i)^{\top}}P \hat x_{i}\right)
\end{equation}
over system \eqref{LTIsys}. 
Moreover, there exists a positive real $\varsigma$ such that \eqref{lyainequ} holds, where $K = {\rm{col}}({K^i})_{i = 1}^N$. 

Choose the following Lyapunov function
\begin{align}
	{V_\epsilon} =& \frac{1}{2}{x^{\top}}Px + \frac{1}{2}\varepsilon _u^{\top}{\left[ {T_u^{\top}({\cal L} \otimes I){T_u}} \right]^{ - 1}}{\varepsilon _u}+ \sum\limits_{i = 1}^N {\frac{1}{{2{\phi^i}}}{{({\beta^i} - {\beta^*})}^2}}  \nonumber \\
	&+ \sum\limits_{i = 1}^N {\frac{1}{{2{\phi _i}}}{{({\gamma _i} - {\gamma ^*})}^2}} + \sum\limits_{i = 1}^N {\frac{1}{{2{\phi _{is}}}}{{({\gamma _{is}} - \gamma_s^*)}^2}}, \label{V_eps} 
\end{align}
where $\beta^*$, ${\gamma^*}$, and $\gamma_s^*$ are three positive constants to be determined later. After substituting \eqref{Busplit} and \eqref{u^iadslidsimp} into \eqref{LTIsys}, differentiating \eqref{V_eps} along the trajectories of \eqref{LTIsys} gives
\begin{align}
	{{\dot V}_\epsilon} =& {x^{\top}}P(A + BK)x + {x^{\top}}PB{K_D}e +\varepsilon _u^{\top}{\left[ {T_u^{\top}({\cal L} \otimes I){T_u}} \right]^{ - 1}}{{\dot \varepsilon }_u} \nonumber\\
	& + \sum\limits_{i = 1}^N {\frac{1}{{{\phi _i}}}({\gamma _i} - {\gamma ^*}){{\dot \gamma }_i}} + \sum\limits_{i = 1}^N {\frac{1}{{{\phi _{is}}}}({\gamma _{is}} - \gamma _s^*){{\dot \gamma }_{is}}}\nonumber\\
	&+x^\top P B_v v - {x^\top}PB{\mathcal B}{{\bar h}_\epsilon} + \sum\limits_{i = 1}^N {\frac{1}{{{\phi^i}}}({\beta^i} - {\beta^*}){{\dot \beta}^i}}, \nonumber
\end{align}
where ${K_D} = {\rm{diag}}({K^i})_{i = 1}^N$, ${{\bar h}_{\epsilon}} = {\rm{col}}\left[h^\iota_{\epsilon}\left({(B^i)^{\top}}P \hat x_{i}\right)\right]_{i=1}^N $, and ${\mathcal B} = {\rm{diag}}({\beta ^i}{I_{{b_i}}})_{i=1}^N$ with $b_i$ denoting the number of the columns in matrix $B^i$.
From \eqref{gagasbeta}, it is obtained that
\begin{subequations}\label{dotVgagasbeta}
	\begin{align}
		\frac{1}{{{\phi _i}}}({\gamma _i} - {\gamma ^*}){{\dot \gamma }_i} =& - \frac{{{\sigma _i}}}{{2{\phi _i}}}\left[ {\gamma _i^2 + {{({\gamma _i} - {\gamma ^*})}^2} - {{({\gamma ^*})}^2}} \right] \nonumber\\
		&+ ({\gamma _i} - {\gamma ^*}){\left\| {{\varepsilon _{iu}}} \right\|^2} \\
		\frac{1}{{{\phi_{is}}}}({\gamma_{is}} - {\gamma_s^*}){{\dot \gamma}_{is}} =& - \frac{{{\sigma_{is}}}}{{2{\phi_{is}}}}\left[ {\gamma_{is}^2 + {{({\gamma_{is}} - {\gamma_s^*})}^2} - {{({\gamma_s^*})}^2}} \right] \nonumber\\
		&+ ({\gamma_{is}} - {\gamma^*}){\left\| {{\varepsilon_{iu}}} \right\|} \\
		\frac{1}{{{\phi^i}}}({\beta^i} - {\beta^*}){{\dot \beta}_i} =& - \frac{{{\sigma^i}}}{{2{\phi^i}}}\left[ {(\beta^i)^2 + {{({\beta^i} - {\beta^*})}^2} - {{({\beta^*})}^2}} \right] \nonumber\\
		&+ ({\beta^i} - {\beta^*}){\left\| {\hat x_i^ \top P{B^i}} \right\|}.
	\end{align}
\end{subequations}
With the relations given in \eqref{dotVgagasbeta}, applying similar treatment as in Section~\ref{prooflfthm} can yield
\begin{align}
	{{\dot V}_\epsilon} \le& -\frac{\varsigma }{8}{x^{\top}}x + \lambda_d \varepsilon _d^{\top}{\varepsilon _d} + ({\lambda_a} + {\lambda_b} + {\lambda_e} - {\gamma ^*})\varepsilon _u^{\top}{\varepsilon _u}- \gamma _s^*\left\| {{\varepsilon _u}} \right\| \nonumber\\
	& -{x^ \top}PB{\mathcal B}{{\bar h}_\epsilon} - \varepsilon _u^ \top T_u^ \top {B_ - }h_{\epsilon}^- +\sum\limits_{i = 1}^N{({\beta^i} - {\beta^*}){\left\| {\hat x_i^ \top P{B^i}} \right\|}}\nonumber\\
	&+x^\top P B_v v - \sum\limits_{i = 1}^N {\frac{{{\sigma _i}}}{{2{\phi _i}}}{{({\gamma _i} - {\gamma ^*})}^2}} + \sum\limits_{i = 1}^N {\frac{{{\sigma _i}}}{{2{\phi _i}}}\left( {{{({\gamma ^*})}^2} - \gamma _i^2} \right)} \nonumber\\
	& - \sum\limits_{i = 1}^N {\frac{{{\sigma_{is}}}}{{2{\phi_{is}}}}{{({\gamma_{is}} - {\gamma_s^*})}^2}} + \sum\limits_{i = 1}^N {\frac{{{\sigma_{is}}}}{{2{\phi_{is}}}}\left( {{{({\gamma_s^*})}^2} - \gamma_{is}^2} \right)} \nonumber\\
	& - \sum\limits_{i = 1}^N {\frac{{{\sigma^i}}}{{2{\phi^i}}}{{({\beta^i} - {\beta^*})}^2}}  + \sum\limits_{i = 1}^N {\frac{{{\sigma^i}}}{{2{\phi^i}}}\left( {{{({\beta^*})}^2} - (\beta^i)^2} \right)}, \label{dotVeps2}
\end{align}
where $B_{-}= {\rm{diag}}(B_{-i})_{i = 1}^N$, and ${h_{\epsilon}^-} = {\rm{col}}(h_{\epsilon}^{-i})_{i = 1}^N$ with $h_{\epsilon}^{-i}$ a vector formed by the sliding mode inputs unavailable to the $i$th observer node. It follows from \eqref{hepsfunc} that
\begin{equation*}
	\left\| {h_{\epsilon}^ - } \right\| \le \left\| {{1_N} \otimes {\left(\mathcal{B}{\bar h}_{\epsilon}\right)}} \right\| = \sqrt N \left\| {{\mathcal{B}{\bar h}_{\epsilon}}} \right\| \le  \sqrt{N \sum\nolimits_{i = 1}^N {{{({\beta ^i})}^2}} } . 
\end{equation*}
Then term $\varepsilon _u^ \top T_u^ \top {B_ - }h_{\epsilon}^-$ in \eqref{dotVeps2} can be estimated as 
\begin{equation}
	\varepsilon _u^ \top T_u^ \top {B_ - }h_\epsilon^ -  \le \frac{N}{\varphi }{\left\| {T_u^ \top {B_ - }} \right\|^2}\varepsilon _u^ \top {\varepsilon _u} + \frac{\varphi }{4}\sum\limits_{i = 1}^N {{{({\beta ^i})}^2}}, 
	\label{normh-eps}
\end{equation}
where $\varphi  = \mathop {\min }\limits_i \frac{{{\sigma ^i}}}{{{\phi ^i}}}$.
Rewrite term ${x^ \top }PB{\cal B}{{\bar h}_\epsilon}$ in \eqref{dotVeps2} as follows:
\begin{equation*}
	- {x^ \top }PB{\cal B}{{\bar h}_\epsilon} =  - {(\hat x - e)^ \top }{P_B}{\cal B}{{\bar h}_\epsilon} =  - \sum\limits_{i = 1}^N {(\hat x_i^ \top  - e_i^ \top )P{B^i}{\beta ^i}h_\epsilon^i}, 
\end{equation*}
where $\hat x = {\rm{col}}({{\hat x}_i})_{i = 1}^N$, ${P_B} = {\rm{diag}}(P{B^i})_{i = 1}^N$,
\begin{align}
	- \hat x_i^ \top P{B^i}{\beta ^i}h_\epsilon^i =& 
	\left\{
	\begin{aligned}
		{ - {\beta ^i}\left\| {\hat x_i^ \top P{B^i}} \right\|,\ \beta ^i}\left\| {\hat x_i^ \top P{B^i}} \right\| &> \epsilon \\
		{ - {\epsilon^{ - 1}}{{({\beta ^i})}^2}{{\left\| {\hat x_i^ \top P{B^i}} \right\|}^2}},\ \beta ^i\left\| {\hat x_i^ \top P{B^i}} \right\| &\le \epsilon
	\end{aligned}
	\right. \nonumber\\
	\le& - {\beta ^i}\left\| {\hat x_i^ \top P{B^i}} \right\| + \frac{\epsilon}{4} \label{hatxPBbhesti}\\
	e_i^ \top P{B^i}{\beta ^i}h_\epsilon^i \le& \frac{\varphi }{4}{({\beta ^i})^2} + \frac{1}{\varphi }{\left\| {e_i^ \top P{B^i}} \right\|^2}.\label{ePBbhesti}
\end{align}
Recall from \eqref{esplitasepsdu} that
\begin{align}
	\frac{1}{\varphi }\sum\limits_{i = 1}^N {{{\left\| {e_i^ \top P{B^i}} \right\|}^2}} =& \frac{1}{\varphi }{\left\| {{e^ \top }{P_B}} \right\|^2}  \nonumber\\
	\le& \frac{2}{\varphi }{\left\| {P_B^ \top {\Delta _d}} \right\|^2}\varepsilon _d^\top {\varepsilon _d} + \frac{2}{\varphi }{\left\| {P_B^ \top {\Delta _u}} \right\|^2}\varepsilon _u^ \top {\varepsilon _u}.\label{phisumePB2}
\end{align}
Next, estimate the following two terms appearing in \eqref{dotVeps2}:
\begin{align}
	&x^\top P B_v v -{\beta^*}\sum\limits_{i = 1}^N{\left\| {\hat x_i^ \top P{B^i}} \right\|} \nonumber\\
	&\le \left\| {{x^ \top }PB} \right\|\left\| {{X_v}v} \right\| - {\beta ^*}\sum\limits_{i = 1}^N {\left( {\left\| {{x^ \top }P{B^i}} \right\| - \left\| {e_i^ \top P{B^i}} \right\|} \right)} \nonumber\\
	&\le \left( {\left\| {{X_v}v} \right\| - {\beta ^*}} \right)\left\| {{x^ \top }PB} \right\| + {\beta ^*}\sqrt N \left\| {{e^ \top }{P_B}} \right\| \nonumber\\
	&\le ({\lambda _f} - {\beta ^*})\left\| {{x^ \top }PB} \right\| + {\beta ^*}\sqrt N \left\| {P_B^ \top {\Delta _d}} \right\|\left\| {{\varepsilon _d}} \right\| \nonumber\\
	& \ \ \ \ + {\beta ^*}\sqrt N \left\| {P_B^ \top {\Delta _u}} \right\|\left\| {{\varepsilon _u}} \right\|, \label{xPBv-bxPB}
\end{align}
where ${\lambda _f} = \left\| {{X_v}} \right\|\bar v$, the first inequality comes from Assumption~\ref{matchcondition}, and the last inequality is from \eqref{esplitasepsdu}.
Combining \eqref{dotVeps2} with \eqref{normh-eps}, \eqref{hatxPBbhesti},
\eqref{ePBbhesti}, \eqref{phisumePB2},  \eqref{xPBv-bxPB}, we can arrive at
\begin{align*}
	{{\dot V}_\epsilon} \le& - \sigma {V_\epsilon} + ({\lambda _f} - {\beta ^*})\left\| {{x^{\top}}PB} \right\| + ({\lambda _g} - {\gamma ^*})\varepsilon _u^{\top}{\varepsilon _u} \nonumber \\
	&+ ({\lambda _h} - \gamma _s^*)\left\| {{\varepsilon _u}} \right\| + {\lambda _k}\varepsilon _d^{\top}{\varepsilon _d} + {\lambda _l}\left\| {{\varepsilon _d}} \right\| + {\lambda_m},
\end{align*}
where
\begin{align*}
	\sigma =& \mathop {\min }\limits_i \left\{ {\frac{\varsigma }{{4\left\| P \right\|}},{\sigma ^i},{\sigma _i},{\sigma _{is}}} \right\},\ {\lambda _h} = {\beta ^*}\sqrt N \left\| {P_B^ \top {\Delta _u}} \right\|\\
	{\lambda _g} =& {\lambda _a} + {\lambda _b} + {\lambda _e} + \sigma \left\| {{{\left[ {T_u^{\top}(\mathcal{L} \otimes I){T_u}} \right]}^{ - 1}}} \right\| \nonumber\\
	&+ \frac{N}{\varphi }{\left\| {T_u^ \top {B_ - }} \right\|^2} + \frac{2}{\varphi }{\left\| {P_B^ \top {\Delta _u}} \right\|^2} \\
	{\lambda _k} =&\lambda_d + \frac{2}{\varphi }{\left\| {P_B^ \top {\Delta _d}} \right\|^2},\ {\lambda _l} = {\beta ^*}\sqrt N \left\| {P_B^ \top {\Delta _d}} \right\| \\
	{\lambda _m} =& \frac{\epsilon{N}}{4} + \sum\limits_{i = 1}^N {\frac{{{\sigma ^i}}}{{2{\phi ^i}}}{{({\beta ^*})}^2}} + \sum\limits_{i = 1}^N {\frac{{{\sigma _{is}}}}{{2{\phi _{is}}}}\left( {{{(\gamma _s^*)}^2} - \gamma _{is}^2} \right)} \\
	& + \sum\limits_{i = 1}^N {\frac{{{\sigma _i}}}{{2{\phi _i}}}\left( {{{({\gamma ^*})}^2} - \gamma _i^2} \right)}.
\end{align*}
Selecting $\beta^* \ge \lambda_f$, $\gamma^* \ge \lambda_g$, and $\gamma_s^* \ge \lambda_h$ guarantees that 
\begin{equation*}
	{{\dot V}_\epsilon} \le - \sigma {V_\epsilon}
	+ {\lambda _k}\varepsilon _d^{\top}{\varepsilon _d} + {\lambda _l}\left\| {{\varepsilon _d}} \right\| + {\lambda_m}.
\end{equation*}
Note that there exist two positive reals $\lambda_n$ and $\sigma_d$ such that 
\begin{equation*}
	{{\lambda _k}\varepsilon _d^{\top}(\tau ){\varepsilon_d}(\tau ) + {\lambda_l}\left\| {{\varepsilon_d}(\tau )} \right\|} \le \lambda_n {\rm{e}}^{-\sigma_d t}.
\end{equation*}
By Lemma~\ref{comparelemma}, it can be obtained that
\begin{align*}
	{V_\epsilon} \le {{\rm{e}}^{ - \sigma t}}{V_\epsilon}(0) + \int_0^t {{{\rm{e}}^{ - \sigma (t - \tau )}}\left( {{\lambda _n}{{\rm{e}}^{ - {\sigma _d}\tau }} + {\lambda _m}(\tau )} \right){\rm{d}}\tau }, 
\end{align*}
where
\begin{align*}
	&\int_0^t {{{\rm{e}}^{ - \sigma (t - \tau )}}\left(\lambda_n {\rm{e}}^{-\sigma_d \tau}\right){\rm{d}}\tau }  \le \frac{{{\lambda _n}({{\rm{e}}^{ - {\sigma _d}t}} - {{\rm{e}}^{ - \sigma t}})}}{{\sigma  - {\sigma _d}}}\\
	&\int_0^t {{{\rm{e}}^{ - \sigma (t - \tau )}}{\lambda _m}(\tau) {\rm{d}}\tau } \\
	&\le \frac{1}{\sigma }
	\left[\frac{\epsilon{N}}{4} + \frac{1}{2}\sum\limits_{i = 1}^N {\left(\frac{{{\sigma ^i}}}{{{\phi ^i}}}{{({\beta ^*})}^2}+{\frac{{{\sigma _{is}}}}{{{\phi _{is}}}} {{{(\gamma _s^*)}^2}} }+{\frac{{{\sigma _i}}}{{{\phi _i}}} {{({\gamma ^*})}^2} } \right)} \right].
\end{align*}
Therefore, it can be concluded that state vector $x$, concatenated state estimation error vector $e$, and adaptive gains exponentially converge to the following set
\begin{equation}\label{residualset}
	\left\{ {\left. {x,e,{\beta ^i},{\gamma _i},{\gamma _{is}}} \right|T_d^ \top e = 0\ {\text{and}}\ {V_o}(x,T_u^ \top e,{\beta ^i},{\gamma _i},{\gamma _{is}}) \le {\lambda _o}} \right\}
\end{equation} 
with a rate at least as fast as $\min \{ \sigma ,{\sigma _d}\}$, where
\begin{align*}
	V_o =& {x^{\top}}Px + {e^ \top }{T_u}T_u^{\top}({\cal L} \otimes I){T_u}T_u^ \top e \nonumber\\
	&+ \sum\limits_{i = 1}^N {\left[ {\frac{1}{{{\phi ^i}}}{{({\beta ^i} - {\beta ^*})}^2} + \frac{1}{{{\phi _i}}}{{({\gamma _i} - {\gamma ^*})}^2} + \frac{1}{{{\phi _{is}}}}{{({\gamma _{is}} - \gamma _s^*)}^2}} \right]} \nonumber\\
	\lambda_o =& \frac{\epsilon{N}}{{2\sigma }} + \frac{1}{\sigma }\sum\limits_{i = 1}^N {\left[ {\frac{{{\sigma ^i}}}{{{\phi ^i}}}{{({\beta ^*})}^2} + \frac{{{\sigma _{is}}}}{{{\phi _{is}}}}{{(\gamma _s^*)}^2} + \frac{{{\sigma _i}}}{{{\phi _i}}}{{({\gamma ^*})}^2}} \right]}.
\end{align*}
By decreasing $\epsilon$ and increasing $\phi^i$, $\phi_{i}$, and $\phi_{is}$, set \eqref{residualset} can be made arbitrarily small, which implies that $\left\|x\right\|$ and $\left\|e\right\|$ can be made arbitrarily close to zero.

\footnotesize
\bibliographystyle{ieeetr}
\bibliography{myref}

\end{document}